# Mechanisms for Making Crowds Truthful

**Radu Jurca**                                                    RADU.JURCA@GMAIL.COM
*Google Inc., Switzerland*
**Boi Faltings**                                                  BOI.FALTINGS@EPFL.CH
*Ecole Polytechnique Fédérale de Lausanne (EPFL)*
*Artificial Intelligence Laboratory (LIA)*
*CH-1015 Lausanne, Switzerland*

## Abstract

We consider schemes for obtaining truthful reports on a common but hidden signal from large groups of rational, self-interested agents. One example are online feedback mechanisms, where users provide observations about the quality of a product or service so that other users can have an accurate idea of what quality they can expect. However, (i) providing such feedback is costly, and (ii) there are many motivations for providing incorrect feedback.

Both problems can be addressed by reward schemes which (i) cover the cost of obtaining and reporting feedback, and (ii) maximize the expected reward of a rational agent who reports truthfully. We address the design of such incentive-compatible rewards for feedback generated in environments with *pure adverse selection*. Here, the correlation between the true knowledge of an agent and her beliefs regarding the likelihoods of reports of other agents can be exploited to make honest reporting a Nash equilibrium.

In this paper we extend existing methods for designing incentive-compatible rewards by also considering collusion. We analyze different scenarios, where, for example, some or all of the agents collude. For each scenario we investigate whether a collusion-resistant, incentive-compatible reward scheme exists, and use *automated mechanism design* to specify an algorithm for deriving an efficient reward mechanism.

## 1. Introduction

An increasing number of applications of artificial intelligence extract knowledge from large groups of agents, also termed the *wisdom of the crowds*. One example are online feedback forums (also known as reputation mechanisms) for obtaining information about the products or services. The testimonies of previous buyers disclose hidden, *experience-related* (Parasuraman, Zeithaml, & Berry, 1985), product attributes such as quality, reliability, ease of use, etc., that can only be observed after the purchase. This previously unavailable information allows the buyers to make better, more efficient decisions, and eliminates some of the problems that would otherwise lead to the collapse of online markets[1].

Recent studies, however, raise important questions regarding the ability of existing reputation mechanisms to reflect the real quality of a product. First, the absence of clear incentives drives only some of the users to voice their opinions. For example, Hu, Pavlou, and Zhang (2006) and Admati and Pfleiderer (2000) show that Amazon[2] ratings of books or CDs follow with great probability bi-modal, U-shaped distributions where most of the ratings are either very good, or very bad. As controlled experiments on the same items reveal normally distributed opinions, the authors conclude that users with a moderate outlook are unlikely to report. Talwar, Jurca, and Faltings (2007) identify another factor that promotes rating, namely the desire to contribute with something new to the previously submitted reports. In both cases, the reputation mechanism collects an unrepresentative sample of reviews that is not necessarily informative for the average user.

---

1. Akerlof (1970) warns about the "Market of Lemons", where asymmetric information drives all, except the worst quality sellers out of the market.

Second, and even more distressful, some users intentionally lie in order to gain external benefits from the distorted reputation. Harmon (2004) reports that some authors write fake reviews on Amazon in order to boost the sale of their own books, or to trash the reputation of competing titles. White (1999) describes manipulation techniques for pushing songs up the charts, and Elliott (2006) and Keates (2007) identify problems associated to fake hotel reviews on the travel reputation site TripAdvisor.com. Although we still see high levels of altruistic (i.e., honest) reporting, the increasing awareness that gains can be made by manipulating online reputation will likely attract more dishonest reporting in the future.

Both problems can be solved if the reputation mechanism rewards users for reporting feedback. First, the reward should cover the cost of reporting, so that more users leave feedback and allow the reputation mechanism to estimate more precisely the quality of products or services. Second, honest feedback should yield higher rewards than lying, so that rational agents find in in their best interest to be truthful. This technique is not limited to reputation mechanisms, but applies more generally to any setting where a private signal is inferred from reports by a crowd of self-interested rational agents.

The reader might already ask whether or not it is reasonable to assume that explicit *payments*[3] will incentivise users to change their reporting behavior. Although humans are known to sometimes act irrationally, there are many examples of online systems where rewards have successfully promoted the elicitation of private information. Prediction markets, for example, consistently outperform traditional prediction tools (Figlewski, 1979; Pennock, Debnath, Glover, & Giles, 2002) and users seem to be equally motivated by fake or real money (Servan-Schreiber, Wolfers, Pennock, & Galebach, 2004). Another example is the ESP Game[4] which extracted an impressive volume of image tags by rewarding players with virtual points. Moreover, the future online economy is likely to contain increasing numbers of automated agents, which by design, will be programmed to behave rationally and maximize their utility.

Fundamental results in the mechanism design literature (d'Aspremont & Grard-Varet, 1979; Crémer & McLean, 1985) show that side payments can be designed to create the incentive for agents to reveal their private opinions truthfully. The best such payment schemes have been constructed based on *proper scoring rules* (Kandori & Matsushima, 1998; Johnson, Pratt, & Zeckhauser, 1990; Clemen, 2002), and exploit the correlation between the observations of different buyers about the same good.

---

2. http://www.amazon.com
3. The term payments is general and includes non-monetary rewards such as preferential access to resources, social status or bonus "points".
4. http://www.espgame.org/

Miller, Resnick, and Zeckhauser (2005) adapt these results to online feedback forums characterized by *pure adverse selection*. In such environments, buyers observe the same innate quality attributes of the products or the service providers, possibly with some noise. The role of the reputation mechanism is *signaling*, i.e. to aggregate reports of the buyers into accurate estimates of these attributes. Examples of such a situation are product rating forums such as Amazon, ePinions or Bizrate, and most services that are provided through machines or networks in an anonymous fashion.

In contrast reputation mechanisms can also be used in a *sanctioning* role to counter *moral hazard*. This exists in environments where the provider can vary the quality attributes for each particular buyer in a strategic manner. The role of the reputation mechanism is to spread information about seller misbehavior and increase its cost to make it unattractive. An example of such an environment are seller ratings in online marketplaces.

The two roles of reputation are complementary, and solve the two important problems associated with online markets (Dellarocas, 2006). The signaling role acts against information asymmetries, and allows agents to accurately identify capable partners. The sanctioning role, on the other hand, acts against cheating incentives and encourages honest behavior.

Like Miller et al. (2005), we concentrate in this paper on pure signaling mechanisms and ignore the effects associated with moral hazard. A set of users is assumed to experience the same product or service, possibly with some noise, and later report the privately perceived quality signal to a central reputation mechanism. The product or service is assumed to have consistent quality over time, so that the quality observed by different users can be modeled as randomly drawn from the same distribution. These assumptions are quite common for services delivered by automated systems like web services or intelligent agents, where failures and degradation in performance are due to random events that have not been planned by a strategic operator.

The reputation mechanism *scores* every submitted feedback by comparing it with another report (called the *reference* report) submitted by a different user about the same good. Miller et al. (2005) prove the existence of general incentive-compatible payments where honest reporting is a Nash equilibrium, and the expected reward is large enough to cover the effort of reporting.

Intuitively, incentive-compatible payments exploit the correlation between the private signal observed by an agent, and the agent's beliefs regarding the reference report. Different quality signals trigger different updates of the agent's private beliefs (by Bayes' Law), and thus modify the agent's expectations regarding the value of the reference report. By paying reporters according to how well their submitted feedback improves the *public* predictor of the reference report (tested against the actual reference report, assumed honest), agents have the incentive to "align" the public predictor to their private beliefs, and thus report the truth. Honest reporting becomes a Nash equilibrium.

Unfortunately, honest reporting is not the only Nash Equilibrium (NE) of such a mechanism. Jurca and Faltings (2005) show that all binary incentive-compatible payment mechanisms using a single reference report have several equilibria; moreover, at least one lying equilibrium gives the agents higher expected payoffs than the truthful NE. This brings forth the problem of collusion, as rational agents could potentially coordinate on a lying equilibrium that gives them a higher payoff than the honest equilibrium.

The simplest such lying equilibrium is for all agents to always report the same, thus leading to perfect prediction of the reference reports. As any product or service in the real world will have occasional defects, truthful reporting will always be a noisy predictor of the reference report, and thus not be able to match the payoff of the lying strategy.

We overcome this problem by using not a single but several reference reports. We start from the observation that in the real world, even the most perfect products or services will occasionally be defective. Thus, we should reward reports that predict this slightly imperfect situation. The key idea is to score a report against a *set* of at least 4 reference reports, and to reward a report according to the distribution of these reference reports, without considering their order. By giving a higher reward for matching all but one of the reference reports, it is possible to give a higher expected payoff to the truthful reporting equilibrium.

As it is difficult to see how to scale the rewards to obtain this characteristic, we use *automated mechanism design* (Conitzer & Sandholm, 2002) to compute the rewards that satisfy the criteria. This technique has been first applied to this problem by Jurca and Faltings (2006) to compute the minimal payments required to ensure honest reporting of reputation information. Jurca and Faltings (2007a) augment the technique by formulating the requirement of collusion safety as additional constraints on the desired mechanism. This paper extends our previous work by presenting a unified framework for designing incentive-compatible, collusion resistant rewards in a broader set of collusion scenarios.

More concretely, we vary the complexity of the collusion scenario along three dimensions. First, we consider the *size of the coalition*, and study what happens when all or only some of the agents can become part of a lying coalition. The *complexity of coordination* is the second dimension, and we consider cases where colluders have or not the necessary sophistication to coordinate on different reporting strategies. Finally, the third dimension addresses *the transfer of utilities* and includes the settings where colluders can make or not side-payments to other colluders.

However, not all combinations are formally treated; some contain contradictory assumptions (e.g., if colluders are assumed capable of side-payments, they should also be assumed capable to coordinate on different strategies) or lead to trivial impossibility results (e.g., collusion resistance is clearly impossible when one strategic agent controls all online identities, exactly the scenario where all agents collude and may transfer payments among themselves).

This paper proceeds as follows. Section 2 formally introduces our model, Section 3 introduces incentive-compatible payment mechanisms and presents some of their properties. Section 4 addresses the design of collusion-resistant reward mechanisms for different scenarios. Finally we discuss related work and future directions to improve our results.

## 2. The Model

We consider an online market where a number of rational buyers (or agents) experience the same product (or service). The quality of the product remains fixed, and defines the product's (unknown) *type*. $\Theta$ is the finite set of possible types, and $\theta$ denotes a member of this set. We assume that all buyers share a common belief regarding the prior probability $Pr[\theta]$, that the product is of type $\theta$. $\sum_{\theta \in \Theta} Pr[\theta] = 1$.

After the purchase, every buyer perceives a binary signal about the quality (i.e., true type) of the product. 1 denotes the high quality signal and captures the satisfaction of the buyer with the product. 0, on the other hand, denotes the low quality signal, or the buyer's dissatisfaction with the product. Every product type is characterized by a different probability distribution over the signals perceived by the buyers. Let $Pr[1|\theta]$ be the probability that the buyer of a product of type $\theta$ is satisfied (i.e., observes the quality signal 1). $Pr[1|\theta_1] \neq Pr[1|\theta_2]$ for all $\theta_1 \neq \theta_2 \in \Theta$, and $Pr[1|\cdot]$ is assumed common knowledge.

To make it simpler for the reader to follow the formal notation, we will present a numerical example. The same example will extended as we introduce new notation, and will serve in the subsequent sections to illustrate our results.

**Example.** *Alice, the owner of an old house, needs some plumbing work done. She knows there are good (type $\theta_G$) and bad (type $\theta_B$) plumbers, such that good plumbers provide high quality service with much higher probability: e.g., $Pr[1|\theta_G] = 0.9$ and $Pr[1|\theta_B] = 0.15$. Alice picks the plumber from the Yellow Pages, and given the reputation of the source, she believes that the plumber, Bob, is likely to be good: e.g., $Pr[\theta_G] = 0.8$ and $Pr[\theta_B] = 0.2$. Therefore, Alice expects to get good service with probability $Pr[\theta_G]Pr[1|\theta_G] + Pr[\theta_B]Pr[1|\theta_B] = 0.75$.*

A central reputation mechanism asks every buyer to submit feedback. Buyers are assumed rational, and not constrained to report the truth. The set of pure reporting strategies of a buyer is $S = \{(s(0), s(1))|s(0), s(1) \in Q_2\}$, where $Q_2 = \{0, 1\}$ is the set of quality signals, and $s = (s(0), s(1))$ denotes the strategy according to which the buyer announces $s(0) \in Q_2$ when she observes low quality, and $s(1) \in Q_2$ when she observes high quality. We will often call the reports 0 and 1 the *negative*, respectively the *positive* report.

To ease the notation, we name the four members of the set $S$ as the honest strategy ($\bar{s}$), the lying strategy ($s^{lie}$), the always reporting one strategy ($s^{pos}$) and the always reporting 0 strategy ($s^{neg}$):

- $\bar{s} = (0, 1)$ the buyer reports 0 when she observes low quality and 1 when she observes high quality;

- $s^{lie} = (1, 0)$ the buyer reports 1 when she observes low quality and 0 when she observes high quality;

- $s^{pos} = (1, 1)$ the buyer reports 1 regardless of her observation;

- $s^{neg} = (0, 0)$ the buyer reports 0 regardless of her observation;

The reputation mechanism rewards buyers for the submitted reports. The payment received by buyer $i$ can depend on any information available to the reputation mechanism: namely, the reports submitted by other buyers, and the common knowledge regarding the environment (probability distribution over types, and conditional probability distributions of quality signals). We assume the reputation mechanism updates the public reputation information with batches of $N$ reports. The agents that submitted the $N$ reports in the same batch are assumed to have had access to the same public information, which motivates the common priors assumption from the beginning of the section. For the rest of the paper will analyze reward mechanisms that work on *static* sets of $N$ reports; in real settings, however, the same mechanisms will be designed over and over again for all batches of size $N$.

Note that the reputation mechanism (i) does not know the true type of the product, and (ii) cannot purchase the product in order to get some first-hand experience regarding its quality.

Discarding from the notation the dependence on the common knowledge, a payment mechanism (employed by the reputation mechanism) is a function $\tau : Q_2 \times (Q_2)^{N-1} \to \mathbb{R}^+$, where $\tau(r_i, r_{-i}) \geq 0$ is the amount paid to buyer $i$ when she reports $r_i \in Q_2$ and the other $N-1$ buyers report $r_{-i} \in (Q_2)^{N-1}$. The reports $r_{-i}$ are also called the *reference reports* of agent $i$, since they constitute the reference for computing the payment for agent $i$. We constrain payments to be non-negative as most online forums cannot impose punishments on the reporters.

The order of reports is not important, therefore we can simplify the payment mechanism by assuming that $\tau(r_i, r_{-i}) = \tau(r_i, r^*_{-i})$ for all $r_{-i}$ and $r^*_{-i}$ that contain the same number of positive reports. A more compact description of the payment mechanism is thus given by the amounts $\tau(r, n)$ where $n \in \{0, 1, \ldots, N-1\}$ is the number of positive reports submitted by the reference reporters.

The payoff expected by agent $i$ depends on the distribution of the reference reports. If the other agents report honestly, the distribution of the reference reports can be computed from the prior beliefs, and the true observation, $o_i \in Q_2$ of agent $i$. The probability that exactly $n$ positive reports were submitted by the other $N-1$ agents is:

$$Pr[n|o_i] = \sum_{\theta \in \Theta} Pr[n|\theta] Pr[\theta|o_i]; \tag{1}$$

where $Pr[n|\theta]$ is the binomial probability distribution function, and $Pr[\theta|o_i]$ can be computed from Bayes' Law:

$$Pr[n|\theta] = \binom{N-1}{n} Pr[1|\theta]^n \left(1 - Pr[1|\theta]\right)^{N-1-n};$$

$$P[\theta|o_i] = \frac{Pr[o_i|\theta] Pr[\theta]}{Pr[o_i]}; \quad Pr[o_i] = \sum_{\theta \in \Theta} Pr[o_i|\theta] Pr[\theta];$$

**Example.** *Once Bob the plumber gets the work done, Alice observes the result and learns something new about Bob's type. If Alice sees good work, her posterior belief regarding the type of Bob will be $Pr[\theta_G|1] = 1 - Pr[\theta_B|1] = 0.96$ (computed by Bayes' Law), and therefore, Alice will believe that some other client will get good service from Bob with probability: $Pr[1|1] = Pr[1|\theta_G] Pr[\theta_G|1] + Pr[1|\theta_B] Pr[\theta_B|1] = 0.87$. On the other hand, if Alice is not happy with the work done by Bob, her posterior belief will be: $Pr[\theta_G|0] = 1 - Pr[\theta_B|0] = 0.32$, and she will expect another client to receive good service from Bob with probability: $Pr[1|0] = Pr[1|\theta_G] Pr[\theta_G|0] + Pr[1|\theta_B] Pr[\theta_B|0] = 0.39$.*

*The reputation mechanism offers Alice the following reward scheme: "the report is paid only if it matches the reference report. A negative report is paid \$2.62, while a positive report is paid \$1.54". Consequently, this reward scheme can be formally described as $\tau(0,0) = 2.62$, $\tau(1,1) = 1.54$, and $\tau(1,0) = \tau(0,1) = 0$. Assuming that the reference report is truthful, one can easily verify that Alice maximizes her expected payment by reporting the truth: If Alice experiences good service from the plumber, she expects that some other client also gets*

*good service with probability 87%. Assuming that the other client reports truthfully, Alice's expected payment is: $.87 \cdot 1.54 + .13 \cdot 0 = 1.34$ if she reports good service, or $.87 \cdot 0 + .13 \cdot 2.62 = 0.34$ if she reports bad service; Likewise, if Alice experiences bad service, she expects that the reference report will be negative with probability $1 - .39 = .61$. In this case, her expected payment is: $.39 \cdot 1.54 + .61 \cdot 0 = 0.6$ if she reports good service, or $.39 \cdot 0 + .61 \cdot 2.62 = 1.6$ if she reports bad service. In both cases, honest reporting is better than lying by $1.*

The numerical example above specifies payments in dollars but does not mention the value of the service that is the object of the reputation report. We specifically avoid the dependence of the reward mechanism on the value of the goods traded in the market. Instead, we will relate the rewards to the *marginal gain for telling the truth*, and the monetary unit is defined as the minimum expected loss of an agent that miss-reports instead of telling the truth.

A strategy profile $s$ is a vector $(s_i)_{i=1,\ldots,N}$, prescribing the reporting strategy $s_i \in S$ for each agent $i$. We will sometimes use the notation $s = (s_i, s_{-i})$, where $s_{-i}$ is the strategy profile for all agents except $i$; i.e., $s_{-i} = (s_j)$, for $j = 1, \ldots, i-1, i+1, \ldots, N$. Given the profile of reporting strategies $(s_i, s_{-i})$, let $\mu[n, s_{-i}]$ describe the belief of agent $i$ regarding the distribution of the reference reports, when:

- $n$ out of the other $N-1$ agents observe the high quality signal, 1

- the other $N-1$ agents are reporting according to the strategy profile $s_{-i}$;

Given $n$ and $s_{-i}$, agent $i$ believes with probability $\mu[n, s_{-i}](x)$ that $x$ reference reports are positive. If $s_i(o_i) \in Q_2$ is the value of the report prescribed by strategy $s_i$ given the true observation $o_i$, the expected payoff to agent $i$ is:

$$V(s_i, s_{-i}|o_i) = \sum_{n=0}^{N-1} Pr[n|o_i] \sum_{x=0}^{N-1} \mu[n, s_{-i}](x) \tau(s_i(o_i), x); \qquad (2)$$

Throughout this paper we will restrict our attention to *pure* reporting strategies and pure strategy equilibria. The reason behind this choice is grounded in practical considerations: mixed strategies and mixed strategy equilibria are more complex and difficult to understand, and therefore unlikely to be observed in practical applications. Acknowledging the limitations brought in by this assumption, we still believe our results are valuable for a number of practical scenarios.

## 3. Incentive-Compatible Payment Mechanisms

In this section we study general payment mechanisms that are incentive-compatible, without worrying about collusion resistance. A payment mechanism is *incentive-compatible* when honest reporting is a Nash Equilibrium (NE): i.e., no agent can gain by lying when other agents report honestly. Formally, let $(\bar{s}_i, \bar{s}_{-i})$ be the strategy profile where all agents report honestly. It is optimal for agent $i$ to report the truth if and only if, for any observation $o_i$, the honest report maximizes the agent's expected payoff:

$$V(\bar{s}_i, \bar{s}_{-i}|o_i) > V(s_i, \bar{s}_{-i}|o_i); \quad \forall s_i \in S \setminus \{\bar{s}\}, o_i \in Q_2;$$

Since reference reports are truthful, the expected payoff to agent $i$ is:

$$V(\bar{s}_i, \bar{s}_{-i}|o_i) = \sum_{n=0}^{N-1} Pr[n|o_i] \tau(o_i, n);$$

and the incentive-compatibility constraints become:

$$\sum_{n=0}^{N-1} Pr[n|o_i]\tau(o_i, n) > \sum_{n=0}^{N-1} Pr[n|o_i]\tau(1-o_i, n); \quad \forall o_i \in Q_2; \quad (3)$$

Practical mechanisms, however, need to offset lying incentives by offering certain margins for truth-telling. Honest reporting must be better than lying by at least some margin $\Lambda$, chosen by the mechanism designer to offset the external benefits an agent might obtain by lying. Rewriting (3) to account for the margin $\Lambda$, an incentive-compatible payment mechanism satisfies the constraints:

$$\sum_{n=0}^{N-1} Pr[n|1]\Big(\tau(1, n) - \tau(0, n)\Big) \geq \Lambda;$$
$$\sum_{n=0}^{N-1} Pr[n|0]\Big(\tau(0, n) - \tau(1, n)\Big) \geq \Lambda; \quad (4)$$

formalizing the intuition that it is more profitable to report positively (respectively negatively) when observing high (respectively low) quality.

Kandori and Matsushima (1998), and Miller et al. (2005) show that it is possible to construct payment mechanisms that satisfy the constraints in (4), based on *scoring rules*. Jurca and Faltings (2006) build on this existence result and describe an algorithm that computes the optimal (i.e., budget minimizing) payment mechanism. We will use this latter approach in this paper, for the obvious practical advantages of designing an incentive compatible reputation mechanism as cheaply as possible.

The expected payment to an honest reporter (in the truthful NE) is the weighted sum between the expected payment to an agent that truthfully reports 1, and the expected payment to an agent that truthfully reports 0:

$$E\Big[V(\bar{s}_i, \bar{s}_{-i})\Big] = Pr[1]\sum_{n=0}^{N-1} Pr[n|1]\tau(1, n) + Pr[0]\sum_{n=0}^{N-1} Pr[n|0]\tau(0, n); \quad (5)$$

where $Pr[1]$ (respectively $Pr[0]$) are the prior probabilities that the agent will perceive high (respectively low) quality, and are defined as: $Pr[o_i] = \sum_{\theta \in \Theta} Pr[o_i|\theta]Pr[\theta]$.

The payment scheme that minimizes the budget required to pay for one honest report therefore solves the linear optimization problem:

**LP 3.1.**

$$min \quad E\Big[V(\bar{s}_i, \bar{s}_{-i})\Big] = Pr[1]\sum_{n=0}^{N-1} Pr[n|1]\tau(1, n) + Pr[0]\sum_{n=0}^{N-1} Pr[n|0]\tau(0, n);$$

$$s.t. \quad \sum_{n=0}^{N-1} Pr[n|1]\Big(\tau(1, n) - \tau(0, n)\Big) \geq \Lambda;$$

$$\sum_{n=0}^{N-1} Pr[n|0]\Big(\tau(0, n) - \tau(1, n)\Big) \geq \Lambda;$$

$$\tau(0, n), \tau(1, n) \geq 0; \forall n = \{0, 1, \ldots, N-1\};$$

Although numerical algorithms can efficiently solve LP 3.1, the analytical solution helps us gain additional insights about the structure of incentive-compatible payment mechanisms. It turns out that LP 3.1 has a simple solution:

**Proposition 3.1.** *The incentive-compatible payment scheme that minimizes the expected payment to an honest reporter (defined by LP 3.1) is:*

$$\tau(0, n) = 0, \forall n \neq 0; \quad \tau(1, n) = 0, \forall n \neq N - 1$$

$$\tau(0, 0) = \Lambda \frac{Pr[N-1|0] + Pr[N-1|1]}{Pr[N-1|1]Pr[0|0] - Pr[N-1|0]Pr[0|1]};$$

$$\tau(1, N-1) = \Lambda \frac{Pr[0|0] + Pr[0|1]}{Pr[N-1|1]Pr[0|0] - Pr[N-1|0]Pr[0|1]};$$

PROOF.

The optimal payment mechanism is symmetric and rewards perfect consensus among reporters: i.e., an agent gets rewarded only if her report agrees with the report of all other agents. The reason why consensus rewards are optimal comes from the structure of the incentive compatible constraints. Clearly there must be at least two positive payments: one rewarding a negative report for some configuration of reference reports, the other rewarding a positive report for some other configuration of reference reports. This proof (the full details are available in Appendix A) shows that it is enough to have only these two positive payments, and that the corresponding configurations of reference reports must reflect consensus.

The first part is intuitively simpler to motivate. The properties of Bayesian updates makes it such that there always exist $n_1$ and $n_2$ such that $Pr[n_1|0] > Pr[n_1|1]$ and $Pr[n_2|1] > Pr[n_2|0]$ (e.g., the configuration where $n_1$ other agents report 1 becomes more probable after a negative experience, while the configuration where $n_2$ other agents report 1 becomes more likely after a positive experience). With potentially infinite payments, the fact that $n_1$ and $n_2$ exist makes it possible to satisfy all incentive compatible constraints; therefore the payment mechanism with only two positive payments for $\tau(0, n_1)$ and $\tau(1, n_2)$ is incentive compatible. The formal proof in Appendix A uses the dual formulation to show the same thing.

The second part of the proof shows that the expected payment is minimized if the scheme rewards consensus (i.e., $n_1 = 0$ and $n_2 = N - 1$). The dual of LP 3.1 reveals that the expected payment to an agent is proportional to the ratios $Pr[n_1|1]/Pr[n_1|0]$ and $Pr[n_2|0]/Pr[n_2|1]$. These ratios reflect the relative change of the agent's beliefs following the subjective private experience. e.g., $Pr[n_1|1]/Pr[n_1|0]$ reflects the relative change for the belief that $n_1$ other agents report 1, given a positive as opposed to a negative experience. Likewise, $Pr[n_2|0]/Pr[n_2|1]$ is the relative change for the belief that $n_2$ other agents report 1, given a negative as opposed to a positive experience. The following lemma shows that these ratios (and therefore the expected payment to an agent) are minimized when $n_1 = 0$ and $n_2 = N - 1$.

**Lemma 3.1.** *Given any set of types $\Theta$, probability distributions $Pr[1|\theta]$, prior belief over types $Pr[\theta]$ and number of agents $N$, we have $\frac{Pr[n|1]}{Pr[n|0]} < \frac{Pr[n+1|1]}{Pr[n+1|0]}$ for all $n = 0 \ldots N - 1$.*

The full proof of the Lemma is also provided in Appendix A. $\qquad\square$

All payment mechanisms[5] that satisfy the incentive compatibility constraints have a similar property: there must be at least two values of the reference reports, $n_1 < n_2$, such that:

$$\tau(0, n_1) > \tau(1, n_2)$$
$$\tau(1, n_2) > \tau(0, n_2);$$

The requirement $n_1 < n_2$ is a direct consequence of the Lemma 3.1. When $\tau(0, n_1)$ and $\tau(1, n_2)$ are scaled appropriately[6], a rational agent prefers the 'bet' on $n_1$ when she observes low quality, and the 'bet' on $n_2$ when she observes high quality.

It is exactly this property that makes it impossible to design an incentive-compatible mechanism that has honest reporting as the unique NE with only one reference report (Jurca & Faltings, 2005). $n_1$ and $n_2$ are constrained to take the values 0, respectively 1, and $\tau(0, 0) > \tau(0, 1)$, $\tau(1, 1) > \tau(1, 0)$, as illustrated by the example in Section 2. Therefore, the constant reporting strategies of always reporting 0 or 1 are also Nash Equilibria. Moreover, since the expected payment to an honest reporter is a linear combination between $\tau(0, 0)$ and $\tau(1, 1)$, at least one of the constant reporting equilibrium generates a higher payoff to the reporters than the honest equilibrium. Hence the vulnerability of the payment mechanism to lying colluders.

Using several reference reports does not, by default, eliminate this problem. The result of Proposition 3.1 shows that the incentive-compatible constraints alone, also generate reward schemes that are vulnerable to conformity rating (i.e, everybody reports the same thing). In most cases, nevertheless, payment schemes based on several reference reports are not *constrained* to reward agreement, so one could specify further conditions, which added to the design problem generate collusion-resistant mechanisms. This is what we will do in the next section. We assume there are $N > 2$ agents in the system and analyze what supplementary constraints can be added to the design problem in order to deter collusion. We consider several collusion scenarios, and whenever possible we present an algorithm that outputs the reward mechanism that is both incentive-compatible and collusion-resistant.

## 4. Collusion-resistant, Incentive-compatible Rewards

The ideal reward mechanism deters any coalition, no matter how big, even when every colluder may use a different strategy and side-payments are possible. Such a mechanism, unfortunately, is trivially impossible: given that all agents may collude and use side-payments to subsidize the agents that might otherwise quit the coalition, the payment mechanism doesn't have any leverage to encourage honest reporting. Whatever the payment scheme, the coalition will adopt the strategy that maximizes the total revenue, regardless of the truth.

Positive results may be obtained only by imposing further restrictions on possible lying coalitions. The first restriction is that not all agents can collude. Some agents are altruistic in nature and report honestly for moral or social reasons. Other agents are not aware of collusion opportunities, or cannot be contacted by a forming coalition. Social or legal norms against collusion may furthermore create prejudices that deter some agents from entering the coalition.

---

5. One might wish, for example, to design a mechanism that minimizes the expected budget paid to all $N$ buyers. In this case, the objective function of the problem LP 3.1 is: $B = \sum_{n=0}^{N} Pr[n](n \cdot \tau(1, n-1) + (N-n) \cdot \tau(0, n))$, where $Pr[n]$ is the prior probability that $n$ out of $N$ buyers observe high quality;

6. $\tau(1, n_1)$ and $\tau(0, n_2)$ will typically be 0

The second restriction addresses the complexity of the coordination among colluders. Symmetric collusion strategies prescribe that all colluders are reporting according to the same strategy. The coordination on symmetric strategies is very simple, and requires one anonymous access to a publicly available source of information that specifies the colluding strategy. Intuitively, the role of the coordination device may be played by a public blog which analyzes the mechanisms and informs potential colluders on the most profitable symmetric colluding strategy. Asymmetric collusion strategies, on the other hand, require significantly more complex coordination. Since every colluder may use a different reporting strategy, the coordination device must know the identity of the colluder before instructing on a collusion strategy. This is often unfeasible, either because colluders might not want to reveal their identity and thus create a trace of their misbehavior, or because identity of the colluders cannot be known at all before the actual reporting takes place.

The third restriction addresses the availability of side-payments between colluders (or transferable utilities). Even when the rewards offered by the reputation mechanism are monetary, the kind of micro-payments that would be required among the colluders are difficult and expensive to implement. Side-payments are even less feasible when the rewards offered by the reputation mechanism are in kind, or in some currency under the control of the reputation mechanism (e.g., Yahoo points or Slashdot karma cannot be transferred even if users wanted to). The conversion of such subjective resources to real money that can afterwards be transferred is even more difficult than the transfer itself.

One notable exception where side-payments are feasible is when the same strategic entity controls a number of online identities, or "sybils" (Cheng & Friedman, 2005). Here, the controlling agent is interested in maximizing his overall revenue (i.e., the sum of the revenues obtained by the sybils), so side-payments do not have to physically occur[7].

To summarize, we address collusion scenarios where:

- all or only some of the agents can become part of a lying coalition,

- colluders can coordinate or not on using different strategies,

- colluders can make or not side-payments to other colluders.

From the remaining seven restricted collusion scenarios (see Table 1) we are only addressing five. We exclude the settings where utilities can be transferred but the coalition is restricted to symmetric strategies. As discussed in the previous paragraph, transferable utilities are mostly characteristic of sybil attacks, where the same strategic agent controls several online identities. We believe it is unreasonable to assume that the strategic agent cannot coordinate the online identities it controls on asymmetric strategy profiles.

For all scenarios involving non-transferable utilities, collusion resistance can emerge as a consequence of having honest reporting as the only (or an attractive enough) equilibrium.

---

7. Whenever rewards are non-monetary, the overall utility of the controlling agent is usually less than the sum of utilities of the sybils. On Slashdot, for example, ten users with bad karma are not worth as one user with good karma. Nevertheless, we will keep for simplicity the assumption of additive utilities for the controlling agent.

| | Non-Transferable Utilities | | Transferable Utilities | |
|---|---|---|---|---|
| | symmetric strategies | asymmetric strategies | symmetric strategies | asymmetric strategies |
| all agents collude | Section 4.1 | Section 4.2 | unreasonable assumption | impossible to prevent collusion |
| some agents collude | Section 4.3 | Section 4.4 | unreasonable assumption | Section 4.5 |

Table 1: Different collusion scenarios.

When all agents may collude, an honest reporting dominant equilibrium is impossible. Therefore, we will resort to designing reward schemes where honest reporting is a unique Nash equilibrium, or a Pareto-optimal Nash equilibrium. When only a fraction of the agents may collude (non-colluders are assumed to report honestly) we also consider designing rewards that make honest reporting the dominant strategy for the colluders. The following subsections address each one collusion scenario, and describe possible methods for designing collusion-resistant, incentive-compatible reward mechanisms.

## 4.1 Full Coalitions on Symmetric Strategies, Non-Transferable Utilities

We assume that agents (i) can only coordinate once (before any of them purchases the product) on the same (pure) reporting strategy, and (ii) cannot make side-payments from one to another. This simple form of coordination between colluders considerably simplifies the problem of the mechanism designer; the only supplementary constraint on the incentive-compatible payment mechanism is to ensure that none of the pure symmetric strategy profiles is a NE.

### 4.1.1 Unique Nash equilibrium.

The set of pure strategies is finite (and contains 3 *lying* strategies) therefore we can exhaustively enumerate the constraints that prevent the corresponding symmetric lying strategy profiles to be NE:

- $s^{pos}$ (always reporting 1) is not NE when a rational agent would rather report 0 instead of 1 given that all other agents follow $s^{pos}$:

$$\tau(0, N-1) > \tau(1, N-1); \tag{6}$$

- $s^{neg}$ (always reporting 0) is not NE when a rational agent would rather report 1 instead of 0 given that all other agents follow $s^{neg}$;

$$\tau(1, 0) > \tau(0, 0); \tag{7}$$

- $s^{lie}$ is not NE when at least one agent (either observing 1 or 0) would rather report the truth. Given that other agents always lie, $N-1-n$ reference reports will be positive whenever $n$ high quality signals were actually observed:

$$\text{either} \sum_{n=0}^{N-1} Pr[n|0]\big(\tau(0, N-1-n) - \tau(1, N-1-n)\big) > 0;$$

$$\text{or} \sum_{n=0}^{N-1} Pr[n|1]\big(\tau(1, N-1-n) - \tau(0, N-1-n)\big) > 0;$$

(8)

The objective function (5), and the constraints (4), (6), (7) and (8) define the optimal incentive-compatible payment mechanism that is also collusion-resistant in the sense explained in the beginning of this section (i.e., honest reporting is the unique pure-strategy symmetric NE). To compute the payments, the mechanism designer must solve two linear optimization problems, one corresponding to each branch of the constraint (8).

**Proposition 4.1.** *Collusion-resistant, incentive-compatible rewards require minimum $N = 4$ agents.*

PROOF. This Proposition is a direct consequence of Proposition 3.1, once we consider the supplementary constraints (7) and (6) that prevent high rewards on unanimous agreement. As we discussed in the proof of Proposition 3.1, any incentive compatible reward mechanism requires two distinct configuration of reference reports – denoted $n_1$ and $n_2$ – such that:

- under the configuration $n_1$ (i.e., $n_1$ reference reports are positive) an agent reporting 0 is rewarded more than an agent reporting 1: e.g., $\tau(0, n_1) > \tau(1, n_1)$

- under the configuration $n_2$ (i.e., $n_2$ reference reports are positive) an agent reporting 1 is rewarded more than an agent reporting 0: e.g., $\tau(1, n_2) > \tau(0, n_2)$

- $n_1 < n_2$ (proven by Lemma 3.1)

The collusion on the positive report (all agents report 1) can be prevented only if $n_2 \neq N-1$, because otherwise $\tau(1, N-1)$ would be greater than $\tau(1, N-1)$, and $s^{pos}$ would be a NE. Likewise, the collusion on the negative report (all agents report 0) can be prevented only if $n_1 \neq 0$ because otherwise $\tau(0,0) > \tau(1,0)$, and $s^{neg}$ would be a NE. Unless $N \geq 4$ the constraints $n_1 \neq 0; n_1 < n_2$ and $n_2 \neq N-1$ cannot be simultaneously satisfied. In other words, there must exist a rich enough set of possible configuration of reference reports in order to satisfy both collusion resistance and incentive compatible constraints. □

Taking the plumber example described in Section 2 and $N = 4$ agents, the conditional distribution of the reference reports can be computed according to Eq. (1):

|   | $Pr[0|\cdot]$ | $Pr[1|\cdot]$ | $Pr[2|\cdot]$ | $Pr[3|\cdot]$ |
|---|---|---|---|---|
| 0 | 0.4179 | 0.2297 | 0.1168 | 0.2356 |
| 1 | 0.0255 | 0.0389 | 0.2356 | 0.7 |

When the experience is negative, the two types ($\theta_G$ and $\theta_B$) become almost equally likely in the agent's private belief. The mix between the distribution of reference reports induced by the two types generates the U-shaped distribution described by the first row of the table. However, when the experience is positive, the good type becomes dominant, and the distribution of reference reports is almost entirely dictated by $Pr[n|\theta_G]$ (the second row of the table). The optimal collusion-resistant, incentive-compatible payment mechanism are the following:

| $\tau(\cdot,\cdot)$ | 0 | 1 | 2 | 3 |
|---|---|---|---|---|
| 0 | 0 | 12.37 | 0 | $\varepsilon$ |
| 1 | $\varepsilon$ | 0 | 6.29 | 0 |

where $\varepsilon$ is a small positive value, and the guaranteed margin for truth-telling is $\Lambda = 1$. For any $N > 4$ the payment mechanism looks the same and rewards a report if all but one of the other agents agree with the submitted report. In the same time, opposing consensus is rewarded by a small amount $\varepsilon$.

Payments with exactly the same structure represent a general solution of the design problem in this context. Moreover, the payments always exist:

**Proposition 4.2.** *Given any set of types $\Theta$, probability distributions $Pr[1|\theta]$, prior belief over types $Pr[\theta]$, and number of agents $N \geq 4$, the following payment system has honest reporting as the unique symmetric NE:*

$$\tau(0,n) = 0, \forall n \neq 1, N-1; \ \tau(1,n) = 0, \forall n \neq 0, N-2; \ \tau(0,N-1) = \tau(1,0) = \varepsilon$$

$$\tau(0,1) = \begin{cases} \Lambda \frac{Pr[1|1]}{Pr[1|0]Pr[1|1]-Pr[N-2|0]Pr[N-2|1]} & \text{if condition A} \\ \Lambda \frac{Pr[1|0]}{Pr[N-2|0]Pr[N-2|1]-Pr[1|0]Pr[1|1]} & \text{if condition B} \\ \Lambda \frac{Pr[N-2|1]+Pr[N-2|0]}{Pr[1|0]Pr[N-2|1]-Pr[N-2|0]Pr[1|1]} & \text{otherwise} \end{cases}$$

$$\tau(1,N-2) = \begin{cases} \Lambda \frac{Pr[N-2|1]}{Pr[1|0]Pr[1|1]-Pr[N-2|0]Pr[N-2|1]} & \text{if condition A} \\ \Lambda \frac{Pr[N-2|0]}{Pr[N-2|0]Pr[N-2|1]-Pr[1|0]Pr[1|1]} & \text{if condition B} \\ \Lambda \frac{Pr[1|1]+Pr[1|0]}{Pr[1|0]Pr[N-2|1]-Pr[N-2|0]Pr[1|1]} & \text{otherwise} \end{cases}$$

$$A = \bigwedge \left( \begin{array}{l} Pr[1|0]Pr[1|1] > Pr[N-2|0]Pr[N-2|1] \\ Pr[N-2|1] > Pr[1|1] \\ Pr[N-2|1]^2 - Pr[1|1]^2 > Pr[1|0]Pr[1|1] - Pr[N-2|0]Pr[N-2|1] \end{array} \right)$$

$$B = \bigwedge \left( \begin{array}{l} Pr[N-2|0]Pr[N-2|1] > Pr[1|0]Pr[1|1] \\ Pr[1|0] > Pr[N-2|0] \\ Pr[1|0]^2 - Pr[N-2|0]^2 > Pr[N-2|0]Pr[N-2|1] - Pr[1|0]Pr[1|1] \end{array} \right)$$

PROOF. It is straight-forward to check that for $\varepsilon$ small enough, the payments described in the proposition verify the constraints (4), (6) and (7). Moreover, these payments minimize the expected payment to an honest reporter. $\square$

### 4.1.2 Pareto-optimal Nash equilibrium.

A less strict notion of collusion resistance requires honest reporting to be a Pareto-optimal NE. The intuition is that any stable (i.e., equilibrium) coalition will necessarily make some colluders worse off than in the honest equilibrium. Assuming non-transferable utilities, colluders that benefit from the coalition cannot subsidize the ones that make a loss, and hopefully, the latter will refuse to join the coalition in the first place. As we will see towards the end of this subsection, payment mechanisms where honesty is a Pareto-optimal NE will be significantly cheaper than the payments designed to have a unique honest NE.

The payment mechanism that has honest reporting as a Pareto-optimal equilibrium solves the following optimization problem:

**LP 4.1.**

$$min \quad E\Big[V(\bar{s}_i, \bar{s}_{-i})\Big] = Pr[1] \sum_{n=0}^{N-1} Pr[n|1]\tau(1,n) + Pr[0] \sum_{n=0}^{N-1} Pr[n|0]\tau(0,n);$$

$$s.t. \quad \sum_{n=0}^{N-1} Pr[n|1]\Big(\tau(1,n) - \tau(0,n)\Big) \geq \Lambda;$$

$$\sum_{n=0}^{N-1} Pr[n|0]\Big(\tau(0,n) - \tau(1,n)\Big) \geq \Lambda;$$

$$\tau(1, N-1) < E\Big[V(\bar{s}_i, \bar{s}_{-i})\Big];$$

$$\tau(0,0) < E\Big[V(\bar{s}_i, \bar{s}_{-i})\Big];$$

$$\bigvee \left( \begin{array}{c} \sum_{n=0}^{N-1} Pr[n|0]\big(\tau(0, N-1-n) - \tau(1, N-1-n)\big) > 0 \\ \sum_{n=0}^{N-1} Pr[n|1]\big(\tau(1, N-1-n) - \tau(0, N-1-n)\big) > 0 \\ E\big[V(s_i^{lie}, s_{-i}^{lie})\big] < E\big[V(\bar{s}_i, \bar{s}_{-i})\big]; \end{array} \right)$$

$$\tau(0,n), \tau(1,n) \geq 0; \forall n = \{0, 1, \dots, N-1\};$$

The first two constraints make honest reporting a Nash equilibrium. The next two constraints prevent lying colluders on $s^{pos}$ or $s^{neg}$ to get higher rewards than in the honest equilibrium. These constraints are always easier to satisfy than the constraints (6), (7) which prevent equilibria on $s^{pos}$ and $s^{neg}$. The last constraint requires that the symmetric profile where every agent lies is either not a NE, or that it generates an expected payoff lower than the honest equilibrium. The expected payoff to an agent reporting according to $s^{lie}$ when everybody else reports according to $s^{lie}$ is:

$$E\big[V(s_i^{lie}, s_{-i}^{lie})\big] = \sum_{n=0}^{N-1} \Big(Pr[n|0]Pr[0]\tau(1, N-1-n) + Pr[n|1]Pr[1]\tau(0, N-1-n)\Big);$$

One remark about LP 4.1 is that it is not always optimal to only consider the constraint that limits the expected payoff of a colluder on $s^{lie}$ below the expected payoff obtained in the honest equilibrium (i.e., the third inequality of the disjunctive constraint of LP 4.1). Numerical simulations performed on random problems show that for 40-50% of the problems the collusion-resistant payments are cheaper by eliminating altogether the symmetric lying equilibrium on $s^{lie}$: i.e., either the first, or the second inequality from the last constraint of LP 4.1 are easier to satisfy than the third inequality.

In either case, the resulting optimal payments have the following structure:

- $\tau(0,0) = \tau(1, N-1) = E\big[V(\bar{s}_i, \bar{s}_{-i})\big] - \varepsilon$. These values prevent the lying coalitions on $s^{pos}$ or $s^{neg}$ to Pareto-dominate the honest reporting equilibrium;

- $\tau(0,1) > 0$ and $\tau(1, N-2) > 0$ are scaled to satisfy the incentive-compatibility constraints, and the "easiest" of the three inequalities that prevent a coalition on $s^{lie}$;

- the other payments are 0.

For the plumber example in Section 2 the payments are the following:

| $\tau(\cdot, \cdot)$ | 0 | 1 | 2 | 3 |
|---|---|---|---|---|
| 0 | 1.30 | 4.52 | 0 | 0 |
| 1 | 0 | 0 | 1.26 | 1.30 |

These payments are much smaller than the ones generated by a mechanism where honest reporting is the unique equilibrium. We therefore observe a fundamental tradeoff between the strongness of collusion resistance guarantees offered by the mechanism, and the price the mechanism has to pay to enforce these guarantees. For the collusion scenario described in this section, Proposition 4.2 shows that it is always possible to design a mechanism where honest reporting is the unique equilibrium. Nevertheless, a mechanism where honest reporting is just Pareto-optimal, not unique, can be significantly cheaper. We will see the same tradeoff in all subsequent scenarios.

Before we move on to the next collusion scenario, let us briefly analyze the influence of $N$ (i.e., the number of reports available to the mechanism) on the properties of the mechanism. Jurca and Faltings (2006) show that the incentive-compatibility constraints are easier to satisfy when $N$ becomes larger. Intuitively, this property is a consequence of the structure of the design problem: the number of constraints in LP 3.1 is independent on $N$, however, the number of variables increases with $N$. Therefore, the dual of LP 3.1 has a constant number of variables, but an increasing number of constraints. Moreover, the constraints of the dual are also harder to satisfy, which means that the maximization objective of the dual can only decrease with $N$. Hence the objective of the primal, i.e., the cost of the mechanism, also decreases with $N$. Without going into the technical details, the same is true for collusion-resistant mechanisms: for larger values of $N$, the collusion-resistance constraints become easier to satisfy and the mechanism has a lower cost. All subsequent sections will maintain this property: more information helps the mechanism designer to specify better targeted rewards, which in turn decreases the total cost of the mechanism.

## 4.2 Full Coalitions on Asymmetric Strategies, Non-Transferable Utilities

The next collusion scenario we are considering is when all $N$ agents can coordinate on asymmetric collusion strategies, without being able to make side-payments from one to another. Each of the $N$ agents can have a different reporting strategy, and the collusion strategy profile is denoted by $s = (s_i)$, $i = 1, \ldots, N$, where $s_i \in S$ is the reporting strategy of agent $i$.

We distinguish between two cases, when the communication (and therefore the coordination on the collusion strategy profile) happens *before* or *after* the agents perceive the quality signals from the product they purchase. In the latter case, no payment scheme can satisfy the incentive-compatibility constraints. In the former case, honest reporting can never be a unique Nash equilibrium of the mechanism; however, honest reporting can be a Pareto-optimal Nash equilibrium.

**Proposition 4.3.** *When agents communicate and coordinate their reports* after *perceiving the quality signals, strict incentive-compatible payment mechanisms do not exist.*

PROOF. Consider two settings, that are identical except for the observation of agent $i$. In setting $I$, agent $i$ observes $o_i = 0$, in setting $II$, agent $i$ observes $o_i = 1$; in both settings the other agents observe $n$ high quality signals. An incentive-compatible mechanism requires $i$ to report 0 in setting $I$, and 1 in setting $II$. Assume all other agents report truthfully; during the communication phase (happening after signals have been perceived) agent $i$ learns in both settings that the reference reports contain $n$ positive reports. An incentive-compatible payment mechanism requires that:

- $\tau(0, n) > \tau(1, n)$ - honest reporting is strictly better for $i$ in setting $I$ ;

- $\tau(1, n) > \tau(0, n)$ - honest reporting is strictly better for $i$ in setting $II$;

Clearly this is impossible. □

The previous proposition formalizes the intuition that truth-telling may only be an ex-ante Nash equilibrium. The reference reports must be unknown to the agent in order to allow the design of incentive-compatible payments.

### 4.2.1 Unique Nash equilibrium.

When the communication takes place before the agents observe the signals, incentive-compatible payments do exist, but always accept other Nash equilibria where agents lie:

**Proposition 4.4.** *When agents communicate and coordinate their reports* before *perceiving the quality signals, no payment mechanism has a unique honest reporting Nash equilibrium.*

Proof. The proof shows that a full coalition can always find a profile of constant reporting strategies, $s = (s_i)$, $i = 1, \ldots, N$, $s_i \in \{s^{neg}, s^{pos}\}$ that is a NE.

We define the family of reporting strategy profiles $s(n) = (s_i)$ where $n$ out of $N$ agents always report 1, and the other $N - n$ agents always report 0: i.e.,

$$s_i = s^{pos}, \ \forall i \in A_1; \quad s_i = s^{neg}, \ \forall i \in A_0;$$
$$|A_1| = n, \quad |A_2| = N - n; \tag{9}$$
$$A_1 \cap A_0 = \varnothing; \quad A_1 \cup A_0 = \{1, 2, \ldots, N\};$$

Assume that the payment mechanism defined by $\tau(\cdot, \cdot)$ accepts honest reporting as the unique NE. We have seen in Section 3 that the incentive-compatible constraints (4) imply the existence of $n_1 < n_2 \in \{0, 1, \ldots, N - 1\}$ such that $\tau(0, n_1) > \tau(1, n_1)$, and $\tau(1, n_2) > \tau(0, n_2)$.

With non-transferable utilities, the strategy profile $s(n_2 + 1)$ is not a NE if and only if one of the $n_2 + 1$ agents that should report 1 would rather report 0:

$$\tau(0, n_2) > \tau(1, n_2);$$

or one of the $N - n_2 - 1$ agents that should report 0 would rather report 1:

$$\tau(1, n_2 + 1) > \tau(0, n_2 + 1);$$

The first inequality cannot be true by the choice of $n_2$; therefore, it must be that $\tau(1, n_2 + 1) > \tau(0, n_2 + 1)$.

Similarly, $s(n_2 + 2)$ is not a NE iff either $\tau(0, n_2 + 1) > \tau(1, n_2 + 1)$ (impossible), or $\tau(1, n_2 + 2) > \tau(0, n_2 + 2)$. Continuing this argument we find that $\tau(1, N - 1) > \tau(0, N - 1)$ which makes $s(N)$ (i.e., all agents report 1) a Nash equilibrium. Hence the result of the proposition. □

Proposition 4.4 holds regardless the number of reports, $N$, available to the reputation mechanism. The proof shows that all incentive-compatible reward schemes have the property that for some $n \in \{0, \ldots, N - 1\}$, either $\tau(1, n) > 0$ and $\tau(1, n + 1) < \tau(0, n + 1)$, or $\tau(1, N-1) > 0$. In the first case, the coalition can adopt the lying strategy where $n+1$ agents always report 1, and $N-n-1$ agents always report 0. The structure of the payments makes such a coalition stable, as no agent finds it profitable to deviate from the coalition. In the second case the payment scheme is vulnerable to everybody always reporting 1.

### 4.2.2 Pareto-optimal Nash equilibrium.

While lying equilibria always exist in this scenario, they do not necessarily *Pareto-dominate* the honest reporting NE. Take for example the incentive-compatible payments that solve LP 3.1, with the additional constraints that $\tau(0,0) = 0$ and $\tau(1, N-1) = 0$. A stable coalition can form on the strategy profiles $s(n_2+1)$ or $s(n_1)$, where $n_2 + 1$ (respectively $n_1$) agents always report 1 and the others always report 0 regardless of their observation. This equilibrium, however, does not Pareto-dominate the truthful one: the agents that report 0 do not get any reward, whereas they do get rewarded in the honest equilibrium.

The payment mechanism can be further improved by setting $\tau(0, n_1-1) = \tau(1, n_2+1) = \varepsilon$, where $\varepsilon$ is some small positive value. This modification eliminates the equilibria $s(n_2+1)$ and $s(n_1)$ and instead introduces the equilibria $s(n_2+2)$ and $s(n_1-1)$. Both these equilibria are extremely unattractive (some agents get paid $\varepsilon$, while others don't get paid at all) and are dominated by the honest equilibrium.

**Proposition 4.5.** *Given the set of types $\Theta$, the conditional probabilities $Pr[1|\theta]$, the prior belief over types $Pr[\theta]$, and $N = 4$ agents, the following payment scheme has honest reporting as a Pareto-optimal Nash equilibrium:*

| $\tau(\cdot, \cdot)$ | 0 | 1 | 2 | 3 |
|---|---|---|---|---|
| 0 | $\varepsilon$ | $x > 0$ | 0 | 0 |
| 1 | 0 | 0 | $y > 0$ | $\varepsilon$ |

*The values $x$ and $y$ depend on the probabilities $Pr[1|\theta]$ and $Pr[\theta]$, and $\varepsilon$ has a small positive value.*

PROOF. The payments here are similar to those of Proposition 4.2 except that consensus is rewarded with some small amount $\varepsilon$ instead of being discouraged. In this way the mechanism has only three NE: honest reporting, always reporting 1 or always reporting 0. Both lying equilibria, however, generate much lower revenues (assuming, of course, that $\varepsilon$ is small enough); therefore, honest reporting is a Pareto-optimal equilibrium. The proof that the mechanism has only 3 NE is based on brute force: for $x$ and $y$ taking the values specified in Proposition 4.2, we verify that no other strategy profile is a NE. The details are presented in Appendix B. □

For general reward mechanisms based on $N > 4$ reports, honest reporting can become a Pareto-optimal NE by considering all lying strategy profiles, $s$, and adding to the design problem either of the following linear constraints:

$$V(s_i, s_{-i}|o_i) < V(s_i^*, s_{-i}|o_i) \text{ for some } i, o_i \text{ and } s_i^*;$$
$$E\big[V(s_i, s_{-i})\big] < E\big[V(\bar{s}_i, \bar{s}_{-i})\big] \text{ for some } i; \tag{10}$$

The first constraint ensures that the strategy profile $s$ is not a NE, and consists of a disjunction of at most 8 linear inequalities: for each reporting strategy $s_i = (s_i(0), s_i(1)) \in S$, the agent reporting according to $s_i$ has the incentive to deviate either when observing 0, or when observing 1. There are four strategies in $S$ and only one possible deviation for each observed signal, hence the 8 inequalities. The second constraint ensures that $s$ does not Pareto-dominate the honest equilibrium, and consists of a similar disjunction of at most 4 inequalities. Note that any two strategy profiles that represent different permutations of the same set of $N$ reporting strategies will generate the same constraints. Therefore, there are $\binom{N+3}{3}$ different constraints imposing honesty as a Pareto-optimal NE, each consisting of a disjunction of at most 12 linear inequations. The resulting optimization problem is a disjunctive linear program which can be transformed into a mixed integer linear program (Sherali & Shetty, 1980).

Unfortunately, the complexity of the resulting optimization problem is exponential in the number $N$ of reporters considered by the payment mechanism. Since the payment mechanism depends on the current belief over the types $\theta$, the reputation mechanism might be required to frequently update the payments in order to reflect the changing beliefs. For large values of $N$ this is clearly infeasible.

We therefore consider a special family of payment mechanisms that can be designed efficiently to make honest reporting a Pareto-optimal NE. The basic idea is to consider payments similar to those from Proposition 4.5, that reward a report only when all but one of the reference reports agree. Consensus on the positive or negative feedback is also rewarded by a small amount $\varepsilon$, but all other payments are zero:

| $\tau(\cdot, \cdot)$ | 0 | 1 | 2...N-3 | N-2 | N-1 |
|---|---|---|---|---|---|
| 0 | $\varepsilon$ | $x > 0$ | 0...0 | 0 | 0 |
| 1 | 0 | 0 | 0...0 | $y > 0$ | $\varepsilon$ |

Figure 1: Payment mechanism for $N > 4$ agents.

The payment mechanism now depends on only 2 parameters, $x$ and $y$ that must be scaled to prevent any other lying strategy profile to become a NE Pareto-dominating the honest equilibrium. Note that no strategy profile where more than one agent reports according to $s^{pos}$ or $s^{neg}$ can become a successful collusion strategy. If at least two agents always report 1, none of the other agents will ever want to report 0 (as $\tau(0, n) = 0$ for any $n \geq 2$). Similarly if at least two agents always report 0, none of the other agents will ever want to report 1. Nevertheless, both consensus equilibria yield very small payoffs, significantly lower than the payoff of the honest reporting equilibrium.

Following the intuition from the proof of Proposition 4.5, many of the remaining lying strategy profiles cannot be a NE regardless of the values of $x$ and $y$. Let us consider the set of potential lying equilibrium strategy profiles:

$$\tilde{S} = \{(n_0 \times s^{neg}, n_1 \times s^{pos}, \bar{n} \times \bar{s}, n_l \times s^{lie}) \mid n_0 + n_1 + \bar{n} + n_l = N\}; \tag{11}$$

where $n_0 \in \{0, 1\}$ agents always report 0, $n_1 \in \{0, 1\}$ agents always report 1, $\bar{n} \notin \{N-1, N\}$ agents report honestly and $n_l$ agents always lie. The cardinality of $\tilde{S}$ is $4(N-1)$. The profile $s \in \tilde{S}$ is a NE if and only if for any strategy $s_i \in s$, the agent reporting according to $s_i$ does not have the incentive to deviate to another reporting strategy given that all other agents keep reporting according to $s_{-i}$. Let $o_i \in Q_2$ be the signal observed by agent $i$. The report prescribed by strategy $s_i$ is $r_i = s_i(o_i) \in Q_2$, and given that $\varepsilon$ is small enough to be ignored, the expected payoff to agent $i$ is:

$$Pr[1|o_i, s_{-i}] \cdot x \text{ if } r_i = 0$$
$$Pr[N-2|o_i, s_{-i}] \cdot y \text{ if } r_i = 1$$

where $Pr[1|o_i, s_{-i}]$ and $Pr[N-2|o_i, s_{-i}]$ are the probabilities that exactly 1, respectively $N-2$ of the other $N-1$ agents will report positively given the observation $o_i$ and the strategy profile $s_{-i}$.

The deviation to reporting $1 - r_i$ is not profitable for some observation $o_i$ if and only if:

$$Pr[1|o_i, s_{-i}] \cdot x - Pr[N-2|o_i, s_{-i}] \cdot y > 0 \text{ if } r_i = 0$$
$$Pr[N-2|o_i, s_{-i}] \cdot y - Pr[1|o_i, s_{-i}] \cdot x > 0 \text{ if } r_i = 1$$

The conditions that make $s$ a NE can therefore be expressed as a set of at most 8 inequalities with the following structure:

$$a_j x - b_j y > 0; \ a_j, b_j > 0$$
$$-a_k x + b_k y > 0; \ a_k, b_k > 0$$

If $\max_j \frac{b_j}{a_j} > \min_k \frac{b_k}{a_k}$ the above system of inequations is infeasible, so that for any positive values of $x$ and $y$ the corresponding strategy profile $s$ can not be a NE. However, when $\max_j \frac{b_j}{a_j} < \min_k \frac{b_k}{a_k}$, there are values of $x$ and $y$ which can make $s$ a NE, and therefore, in the design problem we must specify a constraint that prevents $s$ from Pareto-dominating the honest NE. The corresponding constraint will be a disjunction of inequalities: 2 for restricting $x$ and $y$ to values that do not make $s$ a NE, at most 3 that limit the expected payments of colluders below the expected payment of the honest equilibrium.

Since there are $4(N-1)$ potential lying strategy profiles, the optimization problem defining $x$ and $y$ can have up to $4N-2$ constraints: 2 linear incentive-compatible constraints and up to $4(N-1)$ disjunctive linear constraints. The transformation to a mixed integer linear program involves adding up to $4(N-1)$ integer variables, which in the worst case, can result in exponential-time (in $N$) complexity of the design problem.

Fortunately, we can eliminate most of the strategy profiles in $\tilde{S}$ analytically. It turns out that the payment mechanism from Figure 1 does not accept as a Nash Equilibrium any strategy profile where at lest one agent reports truthfully and another agent reports according to $s^{lie}$:

**Proposition 4.6.** *Let $s = (n_0 \times s^{neg}, n_1 \times s^{pos}, \bar{n} \times \bar{s}, n_l \times s^{lie}) \in \tilde{S}$ be a strategy profile where $n_0$ agents always report 0, $n_1$ agents always report 1, $\bar{n}$ agents report honestly and $n_l$ agents always lie. If $(\bar{n} \neq 0 \wedge n_l \neq 0)$ or $(n_0 = 1 \wedge n_1 = 1)$, $s$ cannot be a Nash equilibrium of the payment mechanism described in Figure 1.*

PROOF. For the reasons explained above, we have restricted the number of agents always reporting 0 or 1 to the following cases: (i) $n_0 = 0, n_1 = 0$, (ii) $n_0 = 1, n_1 = 0$, (iii) $n_0 = 0, n_1 = 1$ and (iv) $n_0 = 1, n_1 = 1$. For all cases, we consider strategy profiles where $\bar{n} \geq 1$ agents are honest, and the remaining agents lie according to $s^{lie}$. For each such profile we show that no values of $x$ and $y$ can simultaneously satisfy the equilibrium constraints of both an honest and a lying agent. Moreover, when $n_0 = n_1 = 1$ we show that the strategy profiles where all other agents are honest or all other agents lie, cannot be Nash equilibria. The technical details of the proof are given in Appendix C. □

The remaining lying strategy profiles to be considered for computing the values of $x$ and $y$ are the following:

- $s_1 = (N \times s^{lie})$ when all agents lie; the constraints to prevent this equilibrium are also considered in LP 4.1;

- $s_2 = \left(s^{neg}, (N-1) \times s^{lie}\right)$ when one agent always reports 0, and all other agents lie;

- $s_3 = \left(s^{pos}, (N-1) \times s^{lie}\right)$ when one agent always reports 1, and all other agents lie;

A solution for $x$ and $y$ can therefore be found in constant time.

## 4.3 Partial Coalitions on Symmetric Strategies, Non-Transferable Utilities

In this section we move our attention from *full* to *partial* coalitions, where not all agents can become part of a lying coalition. The non-colluders are assumed to report honestly, and their reports can be used by the mechanism to deter any "partial" lying coalition. Jurca and Faltings (2005) show that *trusted reports* (reports trusted to be true) are useful in preventing lying coalitions; nevertheless, an important difference from our previous work, is that here, honest reports cannot be identified and selectively used by the reputation mechanism.

### 4.3.1 Unique and Pareto-otpimal Nash equilibrium.

We start as in Section 4.1, by assuming only symmetric collusion strategies and no side-payments available among colluders. The number of colluders is $N_{col} < N$, and the remaining $\bar{N} = N - N_{col}$ report honestly. There are 3 symmetric pure lying strategies, and appropriate constraints can ensure that none of them becomes a Nash equilibrium, or Pareto-dominates the honest equilibrium.

Concretely, let $\bar{Pr}[\cdot|\cdot]$ be the probability distribution of the reports submitted by non-colluders, such that $\bar{Pr}[n|o_i]$ is the probability that $n$ out of $\bar{N}$ agents report positively given the observation $o_i \in Q_2$. Likewise, let $\hat{Pr}[\cdot|\cdot]$ be the probability distribution of the reports submitted by the other colluders: i.e., $\hat{Pr}[n|o_i]$ is the probability that $n$ out of $N_{col} - 1$ colluders report positively.

The payment scheme that makes honest reporting the unique Nash equilibrium for the colluders and minimizes the expected payment to an honest reporter solves the following optimization problem:

$$min \quad E\big[V(\bar{s}_i, \bar{s}_{-i})\big];$$

$$s.t. \quad \sum_{n=0}^{N} Pr[n|0]\big(\tau(0,n) - \tau(1,n)\big) \geq \Lambda;$$

$$\sum_{n=0}^{N} Pr[n|1]\big(\tau(1,n) - \tau(0,n)\big) \geq \Lambda;$$

$$\bigvee_{o_i \in Q_2} \Big(\sum_{n=0}^{\bar{N}} \bar{P}r[n|o_i]\big(\tau(1, n + N_{col} - 1) - \tau(0, n + N_{col} - 1)\big) < 0\Big)$$

$$\bigvee_{o_i \in Q_2} \Big(\sum_{n=0}^{\bar{N}} \bar{P}r[n|o_i]\big(\tau(0,n) - \tau(1,n)\big) < 0\Big)$$

$$\bigvee_{o_i \in Q_2} \Big(\sum_{n=0}^{\bar{N}} \bar{P}r[n|o_i] \sum_{x=0}^{N_{col}-1} \hat{P}r[N_{col} - 1 - x|o_i]\big(\tau(1 - o_i, n + x) - \tau(o_i, n + x)\big) < 0\Big)$$

$$\tau(0,n), \tau(1,n) \geq 0; \forall n = \{0, 1, \ldots, N - 1\};$$

where besides the first two incentive-compatibility constraints, the third, forth and fifth constraints encourage deviations from the symmetric collusion on $s^{pos}$, $s^{neg}$ and $s^{lie}$ respectively. The resulting optimization problem is a disjunctive linear program.

Finally, honest reporting can be made a Pareto-optimal equilibrium by modifying the optimization problem such that the disjunctive constraints preventing an equilibrium on the lying symmetric strategies, also specify inequalities for limiting the payoff received by a colluder below the expected payment of an honest reporter:

- colluders on $s^{pos}$ gain less than in the honest equilibrium:

$$\sum_{n=0}^{\bar{N}} \bar{P}r[n|o_i]\tau(1, n + N_{col} - 1) < E\big[V(\bar{s}_i, \bar{s}_{-i})\big];$$

- colluders on $s^{pos}$ gain less than in the honest equilibrium:

$$\sum_{n=0}^{\bar{N}} \bar{P}r[n|o_i]\tau(0,n) < E\big[V(\bar{s}_i, \bar{s}_{-i})\big];$$

- colluders on $s^{lie}$ expect to gain less than in the honest equilibrium:

$$\sum_{o_i = 0,1} Pr[o_i] \sum_{n=0}^{\bar{N}} \bar{P}r[n|o_i] \sum_{x=0}^{N_{col}-1} \hat{P}r[N_{col} - 1 - x|o_i]\tau(1 - o_i, n + x) < E\big[V(\bar{s}_i, \bar{s}_{-i})\big];$$

Our numerical simulation show that the optimization problem defined above always accepts a solution. We therefore conjecture that it is always possible to design incentive-compatible, collusion resistant rewards under the restrictions discussed in this section. A formal proof of this result remains for future work.

## 4.4 Partial Coalitions on Asymmetric Strategies, Non-Transferable Utilities

As opposed to Section 4.3, in this section we consider a more practical scenario where colluders can also employ *asymmetric* lying strategies: i.e., all strategy profiles $s = (s_i)$, $i = 1, \ldots N_{col}$. Side payments are not allowed among colluders, and we are only concerned with the equilibrium for the $N_{col}$ agents that can become part of the coalition; the remaining $\bar{N} = N - N_{col}$ agents are assumed to report honestly.

### 4.4.1 Unique and Pareto-optimal Nash equilibrium.

The mechanism that makes honest reporting the only Nash equilibrium follows the guidelines derived in Proposition 4.5 for the scenario where all agents can collude: namely the reputation mechanism must consider $N = 4$ reports, the mechanism rewards consensus with a small positive payment $\varepsilon$, and otherwise pays a report if and only if three out of four reports agree. The proof of Proposition 4.5 shows that such a payment scheme accepts three Nash equilibria:

- all agents report honestly,

- all agents always reports 0, and

- all agents always report 1.

.

The restriction made in this section that at least one other does not become part of the coalition (and thus reports the truth) restricts the set of equilibria to only one, where all agents report truthfully. Even when the remaining three agents collude, the only NE under the payment mechanism described by Proposition 4.5 is to report the truth.

General payment mechanisms based on $N > 4$ agents can be designed following the same methodology as in Section 4.2: we will consider all strategy profiles the colluders can use, and add constraints to the design problem such that (i) no lying strategy profile is a NE, or (ii) no NE lying strategy profile Pareto-dominates the honest reporting NE.

Concretely, let $S_{N_{col}}$ be the set of strategy profiles the colluders can use:

$$S_{N_{col}} = \{(n_0 \times s^{neg}, n_1 \times s^{pos}, \bar{n} \times \bar{s}, n_l \times s^{lie}) | n_0 + n_1 + \bar{n} + n_l = N_{col}\}$$

where $s = (n_0 \times s^{neg}, n_1 \times s^{pos}, \bar{n} \times \bar{s}, n_l \times s^{lie}) \in S_{N_{col}}$ is the strategy profile where $n_0$ out of $N_{col}$ colluders always report 0, $n_1$ colluders always report 1, $\bar{n}$ colluders report honestly, and $n_l$ colluders always lie. When colluders report according to the strategy profile $s$, let $(s, \bar{s}) = (n_0 \times s^{neg}, n_1 \times s^{pos}, (\bar{n} + \bar{N}) \times \bar{s}, n_l \times s^{lie})$ be the strategy profile used by the $N$ agents, where the $\bar{N} = N - N_{col}$ non-colluders report honestly.

Honest reporting is the unique Nash equilibrium for colluders if and only if:

$$V\big(s_i, (s_{-i}, \bar{s}) | o_i\big) < V\big(s_i^*, (s_{-i}, \bar{s}) | o_i\big) \tag{12}$$

for some colluder $i$ and observation $o_i$. Similarly, $s$ does not Pareto-dominate the honest reporting equilibrium if:

$$E\Big[V\big(s_i, (s_{-i}, \bar{s})\big)\Big] < E\big[V(\bar{s}_i, \bar{s}_{-i})\big] \tag{13}$$

for some colluder $i$.

Let's compare the constraints expressed above with the similar constraints from Section 4.2 described by Eq. (10). We note two important differences. First, the inequalities apply only to the $N_{col}$ colluders, not to the entire set of agents, so that a strategy profile that is not a NE for the $N$ agents, might still be a Nash equilibrium for colluders. Take, for example, the case where all colluders lie: i.e., $s = (N_{col} \times s^{lie})$. The strategy profile of all agents is therefore $(s, \bar{s}) = (N_{col} \times s^{lie}, \bar{N} \times \bar{s})$. It may well be possible that:

- a lying colluder finds it optimal to report according to $s^{lie}$ given that $\bar{N}$ agents report honestly and $N_{col} - 1$ agents report lies;

- an honest reporter would rather file a negative report after a positive experience, given that $\bar{N} - 1$ agents report honestly and $N_{col}$ agents lie.

So $(s, \bar{s})$ is not a NE when considering all agents, but it is an equilibrium for the subset of colluders. Similarly, it is possible that colluders gain better under $(s, \bar{s})$ and honest reporters gain less, such that $(s, \bar{s})$ Pareto-dominates the honest equilibrium for the colluders, but not for all agents. The constraints (12) and (13) are therefore stricter than their counterparts (10), as non-colluders are assumed to unconditionally report the truth without taking into account the actions of a lying coalition.

Second, we can separately consider the constraint (12), when honest reporting is to be enforced as the unique NE for colluders, or the disjunction of constraints (12) and (13), when honest reporting is to be enforced as a Pareto-optimal NE for colluders. The presence of honest reports makes it possible to design payment mechanisms where honesty is the unique NE, an alternative that was not available under the assumptions of Section 4.2. In both cases, the constraints preventing lying equilibria (or preventing lying equilibria from dominating the honest equilibrium) can be represented by a disjunction of linear inequalities, and consequently, by a conjunction of mixed integer linear constraints. The resulting design problem is a MILP, and, as discussed in Section 4.2, has a worst-time complexity that grows exponentially with the number $N$ of agents.

As in Section 4.2, we can use the payment mechanism from Figure 1 to reduce the complexity of the design problem when honest reporting is the unique or a Pareto-optimal NE. From Proposition 4.6, we know that:

- in any NE, at most one colluder reports according to $s^{neg}$ or $s^{pos}$;

- an honest reporter and a liar cannot both regard their strategies as optimal given that at most one of the other agents reports according to $s^{neg}$ or $s^{pos}$

Therefore, the remaining colluding strategy profiles that must be considered when designing the payments $x$ and $y$ are the following:

- $(N_{col} \times s^{lie})$ when all colluders lie;

- $(s^{neg}, (N_{col} - 1) \times s^{lie})$ when one colluder always reports 0 and the others always lie;

- $(s^{pos}, (N_{col} - 1) \times s^{lie})$ when one colluder always reports 1 and the others always lie;

### 4.4.2 STRONGER EQUILIBRIUM NOTION.

Before evaluating the reward mechanisms that accepts a unique or a Pareto-optimal honest reporting Nash equilibrium, let us note that the subgame restricted to the strategies of the $N_{col}$ colluders accepts a stronger equilibrium concept than Nash equilibrium. When the fraction of colluders is small enough, the available honest reports can make it such that any colluder has the incentive to report honestly no matter what *the other colluders* are reporting. We abuse the notation slightly and call this equilibrium a *dominant* equilibrium. Nevertheless, honest reporting is a dominant strategy only given the $\bar{N}$ honest reporters.

If $\bar{Pr}[\cdot|\cdot]$ describes the probability distribution of the $\bar{N}$ honest reports, and $c$ is the number of positive reports submitted by the other $N_{col} - 1$ colluders, the payments $\tau(\cdot, \cdot)$ that make honest reporting the dominant strategy, and minimize the payment to an honest reporter, are defined by the following optimization problem:

**LP 4.2.**

$$min \quad E\big[V(\bar{s}_i, \bar{s}_{-i})\big] = Pr[1] \sum_{n=0}^{N-1} Pr[n|1]\tau(1,n) + Pr[0] \sum_{n=0}^{N-1} Pr[n|0]\tau(0,n);$$

$$s.t. \quad \sum_{n=0}^{\bar{N}} \bar{Pr}[n|0]\big(\tau(0, n+c) - \tau(1, n+c)\big) \geq \Lambda;$$

$$\sum_{n=0}^{\bar{N}} \bar{Pr}[n|1]\big(\tau(1, n+c) - \tau(0, n+c)\big) \geq \Lambda;$$

$$\forall c \in \{0, \dots N_{col} - 1\},$$

$$\tau(0, n), \tau(1, n) \geq 0; \forall n = \{0, 1, \dots, N-1\};$$

The remaining question is how large may the colluding fraction be, such that collusion-resistant, incentive-compatible mechanisms exist.

**Proposition 4.7.** *When more than half of the agents collude, (i.e., $N_{col} > N/2$), no incentive-compatible payment mechanism can make truth-telling the dominant strategy for the colluders.*

PROOF. The intuition behind the proof is the following: When $N_{col} > N/2$, the $N_{col} - 1$ colluders submit at least as many reports as the remaining $N - N_{col}$ honest reporters. Therefore, any sequence of honest reports, can be 'corrected' by a carefully chosen sequence of colluding reports, such that lying is profitable.

Formally, let us extract from the system of inequalities defined in LP 4.2, the subset corresponding to $c = \{0, \dots, \bar{N}\}$. This subset exists since $\bar{N} < N_{col} - 1$. Let us form the following optimization problem:

$$min \quad Pr[1]\sum_{n=0}^{N-1}Pr[n|1]\tau(1,n) + Pr[0]\sum_{n=0}^{N-1}Pr[n|0]\tau(0,n);$$

$$s.t. \quad \sum_{n=0}^{\bar{N}}\bar{Pr}[n|0]\big(\tau(0,n+c)-\tau(1,n+c)\big) \geq \Lambda;\ \forall c = 0\ldots\bar{N}$$

$$\sum_{n=0}^{\bar{N}}\bar{Pr}[n|1]\big(\tau(1,n+c)-\tau(0,n+c)\big) \geq \Lambda;\ \forall c = 0\ldots\bar{N}$$

$$\tau(0,n),\tau(1,n) \geq 0; \forall n = \{0,1,\ldots,N-1\};$$

Let $y_c^0$ and $y_c^1$ be the dual variables corresponding to the constraints where the colluding agents report $c$ positive signals, and the agent observes 0, respectively 1. The dual problem becomes:

$$max \quad \Lambda\sum_{c=0}^{N_{col}-1}(y_c^0+y_c^1);$$

$$s.t. \quad \sum_{c=0}^{n}\bar{Pr}[n-c|0]y_c^0 - \bar{Pr}[n-c|1]y_c^1 \leq Pr[0]Pr[n|0];\ \forall n = 0\ldots\bar{N}$$

$$\sum_{c=0}^{n}\bar{Pr}[n-c|1]y_c^1 - \bar{Pr}[n-c|0]y_c^0 \leq Pr[1]Pr[n|1];\ \forall n = 0\ldots\bar{N}$$

$$\sum_{c=n+1}^{\bar{N}}\bar{Pr}[\bar{N}+1-c|0]y_c^0 - \bar{Pr}[\bar{N}+1-c|1]y_c^1 \leq Pr[0]Pr[\bar{N}+n+1|0];\ \forall n = 0\ldots\bar{N}$$

$$\sum_{c=n+1}^{\bar{N}}\bar{Pr}[\bar{N}+1-c|1]y_c^1 - \bar{Pr}[\bar{N}+1-c|0]y_c^0 \leq Pr[1]Pr[\bar{N}+n+1|1];\ \forall n = 0\ldots\bar{N}$$

One can easily verify that the dual problem accepts as solutions:

$$\begin{aligned} y_c^1 &= \bar{Pr}[c|0] \cdot const, \\ y_c^0 &= \bar{Pr}[c|1] \cdot const; \end{aligned} \tag{14}$$

for any positive constants. The dual problem is therefore unbounded, which makes the primal infeasible. $\square$

The bound from Proposition 4.7 is also tight. Consider the numerical example from Section 2, and assume the reputation mechanism has $N = 4$ reports. The following payments are resistant to the collusion of $N_{col} = 2$ agents:

| $\tau(\cdot,\cdot)$ | 0 | 1 | 2 | 3 |
|---|---|---|---|---|
| 0 | 1.575 | 3.575 | 0 | 0 |
| 1 | 0 | 0 | 2.203 | 0.943 |

For example, if Alice observes 1, reporting 1 is better than reporting 0 for any report of the other colluder:

$$\bar{Pr}[0|1]\tau(1,0) + \bar{Pr}[1|1]\tau(1,1) + \bar{Pr}[2|1]\tau(1,2) = 1.715;$$
$$\bar{Pr}[0|1]\tau(0,0) + \bar{Pr}[1|1]\tau(0,1) + \bar{Pr}[2|1]\tau(0,2) = 0.715;$$
$$\bar{Pr}[0|1]\tau(1,1) + \bar{Pr}[1|1]\tau(1,2) + \bar{Pr}[2|1]\tau(1,3) = 1.138;$$
$$\bar{Pr}[0|1]\tau(0,1) + \bar{Pr}[1|1]\tau(0,2) + \bar{Pr}[2|1]\tau(0,3) = 0.138;$$

where $\bar{Pr}[0|1] = 0.0385$, $\bar{Pr}[1|1] = 0.1830$ and $\bar{Pr}[2|1] = 0.7785$ are the probabilities that 0, 1, or 2 out of the $\bar{N} = 2$ non-colluders report positively, given that Alice observed high quality.

In the general case, the design problem has $2N_{col}$ different constraints, and therefore, we should expect the budget required by the reputation mechanism to grow with $N_{col}$. We resort to numerical simulations and study the average cost of an incentive-compatible, collusion-resistant reputation mechanism as the fraction of colluders increases. We randomly generated 5000 problems as follows:

- the set of possible types is randomly chosen between 2 and 20;

- for each type, $\theta$, the probability, $Pr[1|\theta]$, that the buyers observe high quality is randomly chosen between 0 and 1;

- we consider reward mechanisms for 5, 10, 15, 20 and 25 agents.

For each problem and every number of agents we varied the number of colluders from 1 to $N/2$. Figure 2 plots the average normalized cost of the collusion-resistant mechanism as a function of the colluding fraction, $N_{col}/N$. One can see that collusion resistance comes almost for free as long as less than one third of the population colludes. Above this bound the cost increases exponentially, which makes most such mechanisms impractical.

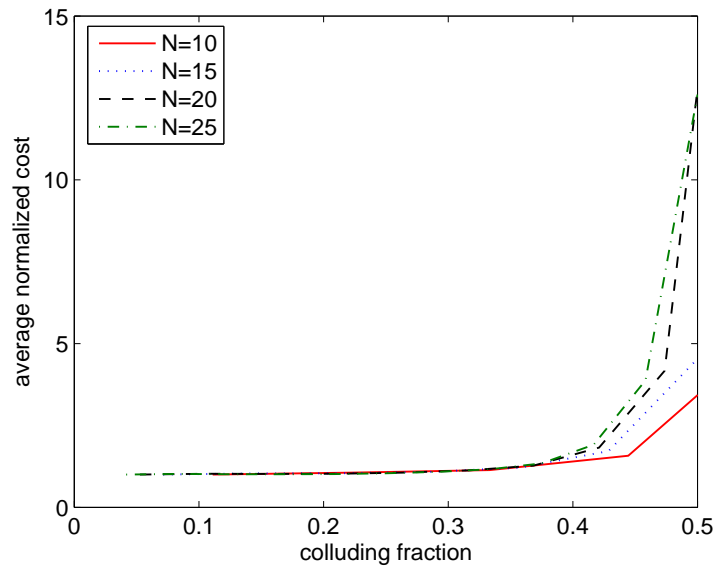

Figure 2: The average cost of the mechanism as we increase the colluding fraction. The cost is normalized to the cost of the corresponding incentive-compatible mechanism that is not collusion-resistant.

Figure 2 seems to contradict the observation made in Section 4.1 that the mechanism has lower cost for higher values of $N$. However, when the same costs are plotted on a horizontal axis describing the *absolute* (not the relative) number of colluders, the order of the lines is reversed: the red line showing the cost of a mechanism with the smallest number of reference reports ($N = 10$) has the highest values.

We also used numerical simulations to investigate the tightness of the bound set by Proposition 4.7. Table 2 presents the distribution of the maximum collusion threshold for the randomly generated problems. For more than 95% of the problems we were actually able to compute payment mechanisms that resist the maximum coalition size $N_{col}^{max} = \lfloor N/2 \rfloor$ described by Proposition 4.7. There are some settings, however, where the mechanism is vulnerable to coalition fractions that are significantly smaller than one half. The sufficient conditions that characterize the settings that accept robust mechanisms with exactly one half of colluders is subject to future research.

Table 2: Distribution of the maximum coalition bound. $N_{col}^{max} = \lfloor N/2 \rfloor$.

| | Distribution of max coalition size (in %) over $[N_{col}^{max}, N_{col}^{max} - 1, \ldots, 1]$ |
|---|---|
| $N = 5, N_{col}^{max} = 2$ | [99.98, 0.02] |
| $N = 10, N_{col}^{max} = 5$ | [99.5, 0.36, 0.1, 0.04, 0] |
| $N = 15, N_{col}^{max} = 7$ | [98.88, 0.54, 0.38, 0.08, 0.1, 0.02, 0] |
| $N = 20, N_{col}^{max} = 10$ | [97.1, 0.86, 0.78, 0.56, 0.34, 0.2, 0.1, 0.04, 0.02, 0] |
| $N = 25, N_{col}^{max} = 12$ | [96.3, 0.98, 0.76, 0.58, 0.48, 0.4, 0.24, 0.1, 0.1, 0.04, 0.02, 0] |

We also compared the performance of the reward mechanisms that employ different equilibrium concepts. Figure 3 compares the average normalized cost of a collusion-resistant payment mechanism when honest reporting is: (i) the dominant strategy, (ii) the unique NE, or (iii) the Pareto-optimal NE. The plots were generated by solving 100 randomly generated problems, for $N = 10$ and $N = 15$ agents. Computing the payment mechanism which satisfies the constraints (12) and (13) requires significantly more time, hence the lower number of generated problems. Moreover, the capabilities of our solver were exceeded for payments using more than 15 agents. Nevertheless, the loss in computational efficiency is clearly rewarded by both lower cost of the mechanism, and coverage of greater coalitions.

## 4.5 Partial Coalitions on Asymmetric Strategies, Transferable Utilities

As a last scenario we assume that one strategic agent controls a number of fake online identities, or *sybils*. From the agent's perspective, the individual revenues obtained by each sybil is irrelevant; the objective of the agent is to maximize the cumulated revenue obtained by all sybils.

The fact that utilities are transferable makes the problem of the mechanism designer significantly harder. In all previous scenarios, the constraints that made an incentive-compatible mechanism collusion-resistant ensured that lying coalitions are either unstable or unprofitable. However, transferable utilities allow some colluders to subsidize others, such that non-equilibrium colluding strategies can still exist. Therefore, the necessary (and sufficient) condition for collusion resistance in this context requires that the cumulated revenue of the coalition is maximized when reporting the truth.

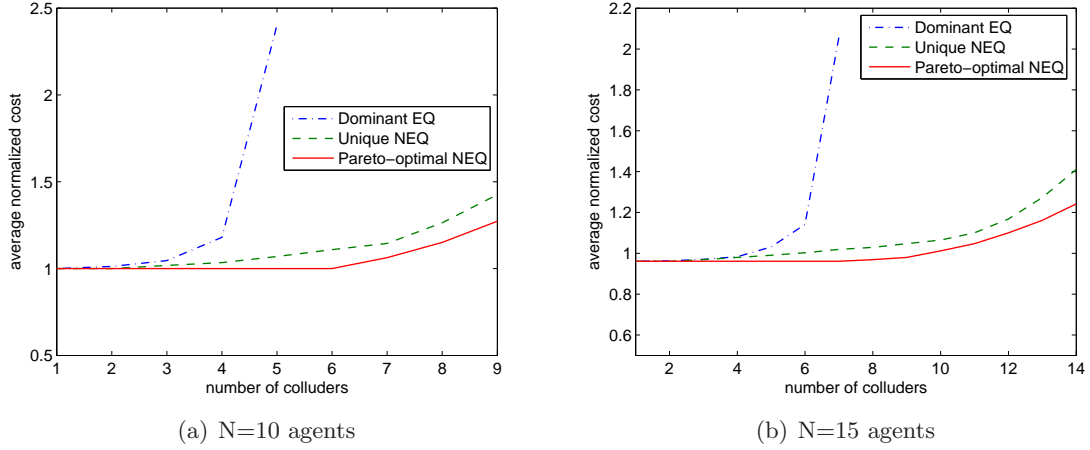

(a) N=10 agents
(b) N=15 agents

Figure 3: Average normalized cost of collusion-resistant payment mechanism. Different equilibrium concepts.

Another difference from the settings in Sections 4.2 and 4.4 is that colluders coordinate their reporting strategy *after* observing the quality signals. This assumption is supported by the interpretation that one strategic entity controls several fake online identities.

Concretely, we are looking for a payment mechanism with the following property: whenever $N_{col}$ colluding agents observe $c$ high quality signals, their cumulated revenue is maximized when reporting $c$ positive reports. An underlying assumption is that non-colluders (the other $\bar{N} = N - N_{col}$ agents) are reporting honestly. The revenue of the coalition that reports $r$ (out of $N_{col}$) can be computed as follows. The $r$ colluders that report positively are rewarded $\tau(1, r-1+n)$, while the $N_{col}-r$ colluders that report negatively are rewarded $\tau(0, r + n)$; $n$ is the number of positive reports submitted by the (honest) non-colluders. The expected revenue of the coalition is therefore:

$$V(r|c) = \sum_{n=0}^{\bar{N}} \bar{P}r[n|c]\Big(r \cdot \tau(1, r-1+n) + (N_{col}-r) \cdot \tau(0, r+n)\Big); \tag{15}$$

where $\bar{P}r[n|c]$ is the probability that $n$ out of $\bar{N}$ honest agents report positively, given that $c$ out of $N_{col}$ colluders observed high quality signals.

Honest reporting is the best strategy for the coalition, when for all $c \in \{0, \ldots N_{col}\}$, $\arg\max_r V(r|c) = c$:

$$\sum_{n=0}^{\bar{N}} \bar{P}r[n|c]\Big(c \cdot \tau(1, c-1+n) + (N_{col}-c) \cdot \tau(0, c+n) - r \cdot \tau(1, r-1+n)$$

$$- (N_{col}-r) \cdot \tau(0, r+n)\Big) \geq \Lambda; \ \forall r \neq c \in \{0, \ldots N_{col}\} \tag{16}$$

The cheapest incentive-compatible, collusion-resistant payment mechanism minimizes the objective function (5) under the linear constraints (16):

**LP 4.3.**

$$min \quad E\Big[V(\bar{s}_i, \bar{s}_{-i})\Big] = Pr[1]\sum_{n=0}^{N-1}Pr[n|1]\tau(1,n) + Pr[0]\sum_{n=0}^{N-1}Pr[n|0]\tau(0,n);$$

$$s.t. \quad (16) \ is \ true, \ \forall c, r \in \{0, \ldots N_{col}\}, c \neq r$$

$$\tau(0,n), \tau(1,n) \geq 0; \forall n = \{0, 1, \ldots, N-1\};$$

For the example described in Section 2, assuming that Alice controls $N_{col} = 3$ different online identities that may submit feedback about Bob, the following payments based on $N = 6$ reports deter Alice from lying:

| $\tau(\cdot,\cdot)$ | 0 | 1 | 2 | 3 | 4 | 5 |
|---|---|---|---|---|---|---|
| 0 | 20.85 | 0 | 0 | 0 | 4.40 | 9.98 |
| 1 | 45.54 | 28.78 | 0 | 0 | 0 | 4.31 |

Even if Alice controlled $N_{col} = 5$ out of the $N = 6$ reports, we can still find payments that make honest reporting rational. These payments, however, are significantly higher:

| $\tau(\cdot,\cdot)$ | 0 | 1 | 2 | 3 | 4 | 5 |
|---|---|---|---|---|---|---|
| 0 | 3455 | 0 | 1378 | 615 | 0 | 1125 |
| 1 | 1530 | 5569 | 4674 | 3736 | 0 | 2585 |

It turns out that for the general case, one honest report is enough to allow the design of incentive-compatible payments that also deter sybil attacks of size $N-1$. An example of such payments are presented in the proposition below:

**Proposition 4.8.** *Given the set of types* $\Theta$, *the conditional probabilities* $Pr[1|\theta]$, *the prior belief over types* $Pr[\theta]$ *and a number* $N$ *of reports, the following payments encourage honest reporting from a strategic agent who controls* $N-1$ *different reports:*

$$\tau(0,0) = \frac{SR(0,0)}{N_{col}}; \ \tau(0,1) = \frac{SR(1,0)}{N_{col}}; \ \tau(1,N) = \frac{SR(1,N-1)}{N_{col}};$$

$$\tau(0,x+1) = \frac{(x+1)SR(1,x) - xSR(0,x+1)}{N_{col}}; \ x = 1\ldots N-1$$

$$\tau(1,x) = \frac{(N-1-x)SR(0,x+1) - (N-2-x)SR(1,x)}{N_{col}}; \ x = 1\ldots N-1$$

*where* $SR(i,j), \ i \in \{0,1\}, \ j = \{0, \ldots, N-1\}$ *is a proper scoring rule: e.g.,* $SR(i,j) = log(\bar{P}r[i|j])$.

PROOF. The expected payment of an agent who controls $N-1$ different identities, observes $c$ out of $N-1$ positive signals and reports $r$ positive reports to the reputation mechanism can be computed as in Eq. (15):

$$V(r|c) = Pr[0|c]\Big(r \cdot \tau(1,r-1) + (N-1-r) \cdot \tau(0,r)\Big) +$$

$$Pr[1|c]\Big(r \cdot \tau(1,r) + (N-1-r) \cdot \tau(0,r+1)\Big)$$

$$= \ldots = Pr[0|c]SR(0,r) + Pr[1|c]SR(1,r);$$

which by the definition of a proper scoring rule is strictly maximized when $r = c$: i.e. $V(c|c) - V(r|c) > 0$ for all $r \neq c$. By scaling the scoring rule appropriately (i.e., multiplication and addition with a constant), honest reporting can be made better by lying with at least the margin $\Lambda$. □

Proposition 4.8 grantees the existence of incentive-compatible collusion resistant rewards when all but one report are controlled by the same strategic agent. However, as we have seen for the example in Section 2, such payments can be very expensive, and hence unpractical. We therefore used numerical simulations to evaluate the marginal cost of increasing collusion resistance, as we increase the number of colluders (i.e., reports controlled by the same agent). As in Section 4.4, we generated 5000 random problems and computed the optimal payments for $N = 5,10,15,20$ and 25 reports. For each case, we gradually increased the coalition size (i.e., $N_{col}$) from 1 to $N - 1$.

Figure 4 plots the average normalized cost of the collusion-resistant mechanism as a function of the coalition fraction. The cost grows exponentially once the coalition fraction covers more than one half of the entire population. This behavior was also observed in Section 4.4 when we evaluated the cost of incentive-compatible, collusion-resistant mechanisms in the absence of transferable utilities (Figure 2). However, the payments required under the assumption of non-transferable utilities are significantly smaller than the payments derived in this section for settings with transferable utilities.

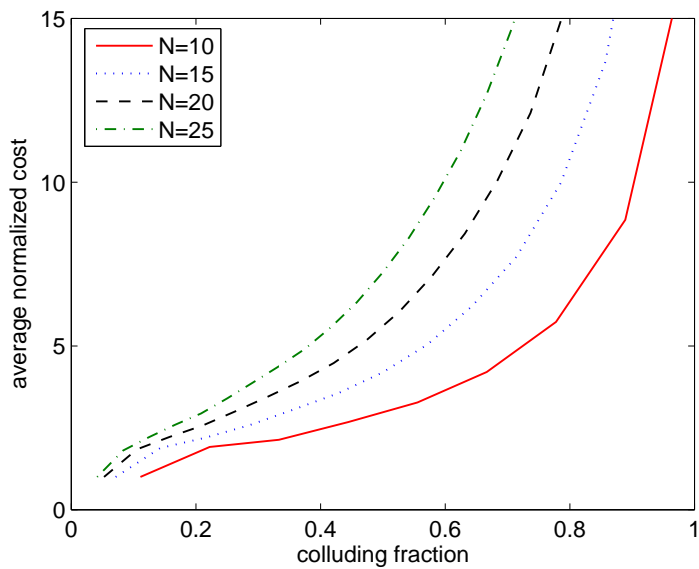

Figure 4: The average cost of the mechanism as we increase the colluding fraction (setting with transferable utilities). The cost is normalized to the cost of the corresponding incentive-compatible mechanism that is not collusion resistant.

The mechanisms defined in the present and the previous sections assume that the mechanism designer knows the total number of colluders. Formally, a mechanism designed to be robust against the collusion of $N_{col}$ agents is not necessarily robust against all coalition sizes smaller than $N_{col}$. Take for example the mechanism defined by Proposition 4.8: a strategic agent that controls less than $N - 1$ identities observes less than $N - 1$ quality signals, and therefore has less precise beliefs about the signals received (and reported) by the non-colluders. These noisy beliefs make it such that in some cases, a smaller coalition may regard a certain lying strategy as more profitable than honest reporting. There is one caveat, however: the colluders know that their lying strategy is provably inefficient had they had access to more information (i.e., the remaining up to $N - 1$ reports). We therefore believe that for all practical purposes, a mechanism designed to be robust against $N_{col}$ colluders will effectively deter all coalitions smaller than $N_{col}$.

## 5. Related Work

One interesting alternative to payment schemes that encourage honest feedback is to develop mechanisms that make it in the best interest of the providers to truthfully reveal their hidden quality attributes. The truthful declaration of quality eliminates the need for reputation mechanisms and significantly reduces the cost of trust management.

Braynov and Sandholm (2002), for example, consider exchanges of goods for money and prove that a market in which agents are trusted to the degree they deserve to be trusted is equally efficient as a market with complete trustworthiness. By scaling the amount of the traded product, the authors prove that it is possible to make it rational for sellers to truthfully declare their trustworthiness. However, the assumptions made about the trading environment (i.e. the form of the cost function and the selling price which is supposed to be smaller than the marginal cost) are not common in most electronic markets.

Another interesting work that addresses the trustworthiness of reputation information is the Goodwill Hunting mechanism by Dellarocas (2002). The mechanism works for eBay-like markets and provides a way to make sellers indifferent between lying or truthfully declaring the quality of the good offered for sale. The particularity of this work is that the goods are advertised to the buyers through the reputation mechanism, which can modify the asking price initially set by the seller. The reputation mechanism thus compensates the momentary gains or losses made by the seller from misstating the quality of the good, and creates an equilibrium where all sellers find it rational to truthfully announce the quality. A major advantage of the mechanism is that it works even when the sellers offer various goods with different values.

Mechanisms for encouraging honest reporting are also present in a number of commercial applications. The most famous is perhaps the *ESP Game* (von Ahn & Dabbish, 2004), designed to encourage human users to label web images. The game[8] pairs two users at random, and shows them the same image. Each player must individually write tags about the image, without being able to see the tags written by the partner. As soon as the two player write the same tag, they gain points and can pass to the next picture. The goal is to get as many points as possible in a fixed amount of time. Intuitively, this game has a very simple strategy: players must write as many correct tags as possible, since the image they see is the only synchronization device that allows them to reach agreement on a tag. The game is very successful, and the authors claim that in this way, all images on the web can be tagged in several months.

---

8. http://www.espgame.org

The incentive mechanism behind the ESP game has, however, several problems. First, it is vulnerable to cheating strategies where a group of players agree to reach agreement on a very simple tag like *"a"* or *"the"*. This strategy could be posted to a popular blog and exposed rapidly to the ESP players. These simple collusion strategies will give colluders a significant competitive advantage, at the detriment of the game designers who collect only garbage tags. The problem can be partly addressed by "taboo" lists containing all confirmed tags which were already submitted about the picture.

The second problem is that rewards are equal for all possible tags. For a Picasso, the players that match on the tag *"painting"* are equally rewarded as the players who correctly identify that the painting is a Picasso. This gives incentives to the players to concentrate on the simplest possible tags like *"person"*, *"man"*, *"woman"*, etc, without spending effort to provide more informative tags. This problem has been partly corrected by the Google Image Labeler[9], a franchise of the ESP Game, which rewards players inversely proportional to the frequency of the tag they agree on. However, the exact algorithm for computing the rewards is not public. Yahoo! is also known to use a version of the ESP Game to tag their collection of images.

Another example of a commercial application using payment mechanisms to encourage honest reporting is Amazon's Mechanical Turk[10]. The role of the system is to provide a marketplace in which human users can solve tasks that are very difficult for machines, but easy for people (i.e., short translations, tagging, face recognition, natural language search, etc). Task owners can pay the workers for answering their tasks, and can also specify payment rules: e.g., a worker gets paid (or receives a bonus) only if the answer is confirmed by a different worker solving the same task.

A number of feedback forums reward raters independently based on the impact of their reviews on the other users. ePinion.com, for example, has professional reviewers who get paid depending on the votes expressed by normal users, and on the purchases made after reading the reviews. Another example is the startup Friend2Friend.com[11] who allows users to gain commissions by recommending products to their friends.

Central to the results of this paper is the principle of automated mechanism design (AMD). The mechanism is created automatically (using optimization algorithms) for the specific problem instance, given the specific information available to the mechanism designer. The idea has important advantages since (a) it can be used to address classes of problems for which there are no known manually designed mechanisms, (b) it can circumvent impossibility results by restricting the mechanism to one particular setting, (c) it can generate better mechanisms by capitalizing on the specific information available in the present setting, and (d) it shifts the effort of mechanism design to a machine.

Since first introduced by Conitzer and Sandholm (2002), AMD was used to generate several impressive results. Conitzer and Sandholm (2003a) (a) reinvented the Mayerson auction which maximizes the seller's expected revenue in a single-object auction, (b) created expected revenue maximizing combinatorial auctions, and (c) created optimal mechanisms for a public good problem. Guo and Conitzer (2007) use AMD to optimally redistribute the payments generated by the VCG mechanism, Conitzer and Sandholm (2007) incrementally design incentive compatible mechanisms, while Hajiaghayi, Kleinberg, and Sandholm (2007) focus on AMD for online settings. Conitzer and Sandholm (2003b) show that the AMD can potentially be exponentially faster for settings with structured preferences that allow a concise representation of the input. Conitzer and Sandholm (2004) describe an efficient algorithm for AMD when the mechanism is deterministic, does not allow payments and there is only one type-reporting agent. AMD can also be used to design multi-stage mechanisms that reduce the burden of information elicitation by querying the agents only for relevant information (Sandholm, Conitzer, & Boutilier, 2007). The results of this paper add to this already long list of results obtained through AMD.

The results of Section 4 are mostly related to the literature on implementation theory and incentive contracts for principle-(multi)agent settings. The main goal of implementation theory is to characterize the space of social choice rules that are implementable by some mechanism given a game-theoretic equilibrium concept. For complete information settings, well established results characterize the necessary and sufficient conditions for a social choice rule (SCR) to be implementable in dominant strategy or in Nash equilibrium. For example, SCRs can be implemented in dominant strategies only if they are strategy-proof (Gibbard, 1973), while the SCRs that can be Nash-implemented must satisfy the property of *monotonicity* and *no veto power* (Maskin, 1999). Unfortunately, SCRs of practical interest do not satisfy the monotonicity requirement. Fortunately, non-monotonic SCRs can be implemented in undominated Nash equilibria (Palfrey & Srivastava, 1991), or in subgame perfect equilibria by multi-stage mechanisms. Another relaxation that extends the set of implementable SCRs is to consider *virtual* implementation, where the socially optimal outcome is required to occur only with probability close to one (Matsushima, 1988; Abreu & Sen, 1991).

In environments with incomplete information agents have private information that is not shared by other agents. The truthful revelation of the private information can only be ensured by social choice rules that are Bayesian incentive-compatible. Moreover, a Bayesian monotonicity condition is necessary for Bayesian implementation (Jackson, 1991). Moore and Repullo (2005) characterize SCRs that can be *virtually* Bayesian implemented in pure strategies, and derive the necessary and sufficient conditions of incentive compatibility and *virtual monotonicity.*

However, applying the implementation theory to the feedback reporting setting (an environment with incomplete information) provides nothing more than the constraints on the payment function such that honest reporting is the unique Bayesian Nash equilibrium. In implementation theory terms, the set of possible world states consists of all combinations of $N$ privately perceived quality signal (one signal for each agent). The outcome space contains all possible sets of $N$ feedback reports and all possible combinations of $N$ positive payments made to the $N$ agents. The desirable SCR contains all social choice functions that map the possible states of the world (i.e., the set of privately perceived signals) to the outcomes where the reported feedback correspond to the privately perceived signals). Implementation theory tells that the SCR must be incentive compatible (i.e., the social choice functions prescribe outcomes where the payments to the agents make them truthfully reveal their private information) and Bayesian monotone (i.e., the social choice functions prescribe outcomes where the payments received by the agents make honest reporting the unique equilibrium). The results of Section 4 translate these requirements into practical constraints that allow the computation of payment functions (and therefore social choice functions) that are Bayesian Nash implementable.

A number of papers discuss incentive contracts that a principal should offer to several agents whose effort levels are private. The reward received by each agent depends on the output observed by the principal, and on the declarations of other agents. Holmström (1982), Ma (1988), and Li and Balachandran (2000) show that efficient contracts exist that are also incentive-compatible and collusion-proof. While the feedback reporting problem is similar, it differs in one major aspect: the reputation mechanism designer (i.e., the principal) does not observe a direct signal which is correlated to the reporters' (i.e., agents') private information.

## 6. Discussion and Future Work

Throughout this paper we only considered pure reporting strategies. Extending our results to *mixed*-strategy equilibria remains an open question, and poses non-trivial computational problems: the constraints required to prevent mixed equilibria are no longer linear, and this significantly increases the complexity of the design problem.

In this paper we limit our investigation to binary settings where the quality signal observed by the agents and the feedback reported to the reputation mechanism is either 0 or 1. While the extension to n-ary feedback is conceptually straight-forward (see Appendix D), the resulting mechanism design problem becomes exponentially more complex as we increase the number of possible feedback values. It remains a challenge for future work to design efficient algorithms that are able to quickly compute incentive-compatible and collusion-resistant reward mechanisms for non-binary feedback.

Another challenge is to relax the requirement of common prior information. The model we introduce in Section 2 assumes that all agents share the same prior beliefs regarding the probability of positive or negative feedback. Jurca and Faltings (2007b) show that incentive compatible reward schemes can still be designed when agents have small amounts of private information. However, the same methodology cannot be easily extended to also address collusion.

Yet another direction for future research is to design payment mechanisms that are resistant against more complex collusion scenarios, where, for example, several strategic agents, each controlling several fake identities, try to manipulate the reporting mechanism.

We believe, however, that there are many practical scenarios where the techniques we presented in this paper can be successfully used to ensure safety against collusion. One example is the monitoring of service quality. Jurca, Binder, and Faltings (2007) describe a framework for monitoring the quality of web service based on client feedback. The idea is to estimate the quality delivered by a service provider directly from the reports of the clients. This approach is much cheaper and more precise when compared to traditional monitoring devices that proxy and analyze the communication between the client and the server. Nevertheless, the mechanism must provide incentives for honest reporting which also discourage lying coalitions. The feedback in such settings is often binary, and specifies whether or not a certain service level agreements (SLA) has been met by the provider. The quality of a provider is than defined by its capacity to fulfill the SLA, and usually becomes public information. The quality at time $t$ defines the priors for designing the mechanism at time $t + 1$, and it is quite natural that the majority of the clients will consider these priors to be public knowledge.

Many tasks on Amazon's Mechanical Turk also fit the model we used in this paper. For example, certain tasks ask human raters to specify whether two descriptions refer to the same item or not. In other tasks, raters must vote thumbs up or down for a piece of news, or must tag photographs that contain a certain visual clue, such as a human face. The answer to all these tasks can be modeled by a binary feedback signal (e.g., items are the same or not, positive or negative vote, a photograph contains or not a human face) and the answers of different users may be considered as conditionally independent given the description of the task. Moreover, these conditional probabilities may quite often be assumed common knowledge: for example, the assumption that 98% of the raters will correctly identify a human face in a decent quality photograph, is not only natural, but is also likely to be a belief of most internet-savvy humans.

More generally, we believe our techniques can be useful in providing incentives for human raters to label training data for supervised machine learning algorithms. Many practical classifiers are binary (e.g., a photograph contains or not a certain feature, a word is misspelled or not) or composed of other binary classifiers (e.g., recognizing a hand-written digit). A framework similar to the Mechanical Turk can harvest the power of the crowds to produce extensive training sets for different algorithms.

Another potential application can be the collaborative question answering forums like Yahoo! Answers. In such forums users may post questions that remain open for a predefined period of time. During this period, other users can write new answers, or vote for existing answers. The voting mechanism is essential for differentiating good from bad answers, and proper incentives may ensure higher participation and more accurate results. These forums are especially vulnerable to collusion, since the author providing the best answer is often rewarded by the forum with points, public recognition, or other benefits. The biggest challenge in such contexts is to obtain accurate estimates on the different probabilities that enter the design problem. For example, the prior probability of a high quality answer can be estimated from the history of the site. On the other hand, the conditional probability that a user will find a given answer as high quality is more complicated to estimate. For example, the mechanism designer might use a natural language processing algorithm to figure out a degree of matching between the question and the answer. The designer could also search for documents that contain keywords related to the question and to the answer, and analyze these documents to refine the matching degree of an answer to the given question. General user statistics can also be factored in to estimate the likelihood that a random user will find a given answer useful. Although these estimates will inevitably be noisy, they might work well enough for the average user.

If our results can be fully extended to feedback sets of arbitrary size, the techniques of this paper will be relevant for most feedback reporting scenarios. To mention just one supplementary examples, the ESP game, where every photo tag can be assumed to be drawn from a finite set of concepts, and the conditional probabilities of seeing a certain tag can be estimated by using word frequencies in different languages, image recognition techniques, and historical data regarding the distribution of tags in similar photos.

| | Non-Transferable Utilities | | Transferable Utilities | |
| --- | --- | --- | --- | --- |
| | symmetric strategies | asymmetric strategies | symmetric strategies | asymmetric strategies |
| all agents collude | -unique honest NE; -Pareto-optimal honest NE | -Pareto-optimal honest NE | unreasonable assumption | impossible to prevent collusion |
| some agents collude | -unique honest NE; -Pareto-optimal honest NE | -unique honest NE; -Pareto-optimal honest NE; -sometimes honest dominant strategy (not if $N_{col} \geq \frac{N}{2}$); | unreasonable assumption | -(sybil attack), the coalition maximizes its revenue by reporting honestly; |

Table 3: Summary of results.

## 7. Conclusion

As online feedback and reputation become increasingly important sources of information, explicit measures must guarantee that honest reporting is in the best interest of the participants. Previous work shows that it is possible to construct payment mechanisms that make honest reporting a Nash equilibrium, because agents expect to get rewarded more for a truthful report than for a lie. Unfortunately, such mechanisms also have other equilibria where reporters lie. This creates collusion opportunities, since several agents can coordinate their lies in order to improve their revenues.

In this paper we addressed the design of incentive-compatible payments that are also resistant to collusion. We consider different collusion scenarios where (i) all or only some of the agents collude, (ii) colluders can coordinate on symmetric or asymmetric strategy profiles, and (iii) colluders can transfer or not payments to each other. Table 3 summarizes the results obtained for each scenario.

In Section 4.1 we assume that all agents may collude but cannot make side-payments to each other. We showed that incentive-compatible payments can be efficiently constructed such that honest reporting is the unique pure strategy symmetric NE, or a Pareto-optimal pure symmetric NE. In Section 4.2 we keep the same assumptions, but investigate asymmetric collusion strategies. We find that any incentive-compatible payment mechanism also accepts asymmetric lying equilibria. Nevertheless, there are payment mechanisms where honest reporting is a Pareto-optimal NE.

Sections 4.3 and 4.4 assume that only a fraction of the agents may collude, and that non-colluders report honestly. If colluders can only coordinate on symmetric strategy profiles and they cannot make side-payments to each other, payments always exist such that honest reporting can be made the unique NE or (with lower payments) a Pareto-optimal NE. If colluders can coordinate on asymmetric strategies (Section 4.4), payments can still be devised that make honest reporting the unique or Pareto-optimal NE. If less than one half the population can collude, then payments can sometimes be devised that make honest reporting a dominant strategy. Numerical simulations, however, show that the payments required to deter coalition fractions greater than one third become exponentially expensive.

Finally, Section 4.5 describes incentive-compatible payments that are resistant to *sybil* attacks: i.e., the same strategic agents creates several fake identities in order to manipulate the payment mechanism. The designer can ensure that the set of reports submitted by the coalition reflects the aggregated experience of the coalitions. Individual colluders do not necessarily report the truth, but overall, the reputation mechanism obtains correct information.

## Acknowledgments

The authors wish to thank the anonymous reviewers for their helpful comments and suggestions.

## Appendix A. Proof of Proposition 3.1

For solving LP 3.1, let us write the corresponding dual problem:

$$
\begin{aligned}
max \quad & \Lambda y_0 + \Lambda y_1; \\
s.t. \quad & Pr[n|0]y_0 - Pr[n|1]y_1 \le Pr[0]Pr[n|0] \\
& Pr[n|1]y_1 - Pr[n|0]y_0 \le Pr[1]Pr[n|1] \\
& \forall n \in \{0, \dots, N-1\};
\end{aligned}
$$

where $y_0$ (respectively $y_1$) is the dual variable corresponding to the constraint where the agent observes 0 (respectively 1). By dividing the first set of constraints by $Pr[n|0]$ and the second set of constraints by $Pr[n|1]$, we have:

$$
\begin{aligned}
y_0 - y_1 Pr[n|1]/Pr[n|0] \le Pr[0], \forall n \in \{0, \dots, N-1\}; \\
y_1 - y_0 Pr[n|0]/Pr[n|1] \le Pr[1], \forall n \in \{0, \dots, N-1\};
\end{aligned}
$$

Clearly, among the $2(N-1)$ constraints of the dual problem, only two are active, corresponding to: $n_1 = \arg\min_n \frac{Pr[n|1]}{Pr[n|0]}$, and $n_2 = \arg\min_n \frac{Pr[n|0]}{Pr[n|1]}$. It follows that only two of the variables of LP 3.1 have non-zero values (i.e., $\tau(0, n_1) \ne 0$ and $\tau(1, n_2) \ne 0$), and they satisfy the linear equations:

$$
\begin{aligned}
Pr[n_1|0]\tau(0, n_1) - Pr[n_2|0]\tau(1, n_2) = \Lambda; \\
-Pr[n_1|1]\tau(0, n_1) + Pr[n_2|1]\tau(1, n_2) = \Lambda;
\end{aligned}
$$

The remaining part of the proof is to show that $n_1 = 0$ and $n_2 = N-1$. For that, we will prove that $\frac{Pr[n|1]}{Pr[n|0]} < \frac{Pr[n+1|1]}{Pr[n+1|0]}$ for all $n = 0, 1, \dots, N-2$.

$$Pr[n|1]Pr[n+1|0] - Pr[n|0]Pr[n+1|1] = \Big( \sum_{\theta \in \Theta} Pr[\theta] \frac{Pr[1|\theta]}{Pr[1]} Pr[n|\theta] \Big) \Big( \sum_{\theta \in \Theta} Pr[\theta] \frac{Pr[0|\theta]}{Pr[0]} Pr[n+1|\theta] \Big) -$$

$$\Big( \sum_{\theta \in \Theta} Pr[\theta] \frac{Pr[0|\theta]}{Pr[0]} Pr[n|\theta] \Big) \Big( \sum_{\theta \in \Theta} Pr[\theta] \frac{Pr[1|\theta]}{Pr[1]} Pr[n+1|\theta] \Big)$$

$$= \Big( \sum_{\theta \in \Theta} Pr[\theta] \frac{Pr[1|\theta]}{Pr[1]} Pr[n|\theta] \Big) \Big( \sum_{\theta \in \Theta} Pr[\theta] \frac{Pr[0|\theta]}{Pr[0]} Pr[n|\theta] \frac{(N-1-n)Pr[1|\theta]}{(n+1)Pr[0|\theta]} \Big) -$$

$$\Big( \sum_{\theta \in \Theta} Pr[\theta] \frac{Pr[0|\theta]}{Pr[0]} Pr[n|\theta] \Big) \Big( \sum_{\theta \in \Theta} Pr[\theta] \frac{Pr[1|\theta]}{Pr[1]} Pr[n|\theta] \frac{(N-1-n)Pr[1|\theta]}{(n+1)Pr[0|\theta]} \Big)$$

$$= \frac{(N-1-n)}{(n+1)Pr[1]Pr[0]} \Bigg( \Big( \sum_{\theta \in \Theta} Pr[\theta] Pr[1|\theta] Pr[n|\theta] \Big)^2 -$$

$$\Big( \sum_{\theta \in \Theta} Pr[\theta] Pr[0|\theta] Pr[n|\theta] \Big) \Big( \sum_{\theta \in \Theta} Pr[\theta] \frac{Pr[1|\theta]^2}{Pr[0|\theta]} Pr[n|\theta] \Big) \Bigg) < 0;$$

by the Cauchy-Schwartz inequality applied to the vectors $(\sqrt{Pr[\theta]Pr[0|\theta]Pr[n|\theta]})_{\theta \in \Theta}$ and $(\frac{Pr[1|\theta]\sqrt{Pr[\theta]Pr[n|\theta]}}{\sqrt{Pr[0|\theta]}})_{\theta \in \Theta}$.

## Appendix B. Proof of Proposition 4.5

The idea of the proof is to show that we can find the positive values $x$ and $y$ such that the payment scheme defined in Proposition 4.5 has only three NE: honest reporting, everybody reporting 0 or everybody reporting 1.

*No NE where $n$ agents report according to $s^{lie}$ and $4 - n$ agents report honestly.* From Proposition 4.2 we know that $x$ and $y$ can be found to prevent an equilibrium where all agents lie. Similarly, the incentive-compatible constraints ensure that a strategy profile where one agent always lies and three agents always report the truth cannot be a NE. Let us show that the profile $s = (3 \times s^{lie}, \bar{s})$ where three agents lie and one agent reports the truth. The honest reporter observing a low quality signal will report honestly if and only if:

$$Pr[2|0]x - Pr[1|0]y > 0;$$

The same honest agent reports a positive report after observing high quality if and only if:

$$-Pr[2|1]x + Pr[1|1]y > 0;$$

However, by Lemma 3.1 we have $\frac{Pr[1|0]}{Pr[2|0]} > \frac{Pr[1|1]}{Pr[2|1]}$, so the two inequalities can never be simultaneously satisfied.

Consider the profile $s = (2 \times s^{lie}, 2 \times \bar{s})$ where two agents lie and two agents report the truth is not NE. One honest reporter reports the truth if and only if:

$$(3Pr[3|0] + 2Pr[1|0])x - (3Pr[0|0] + 2Pr[2|0])y > 0;$$
$$-(3Pr[3|1] + 2Pr[1|1])x + (3Pr[0|1] + 2Pr[2|1])y > 0;$$

A liar, on the other hand, reports according to $s^{lie}$ if and only if:

$$(3Pr[0|1] + 2Pr[2|1])x - (3Pr[3|1] + 2Pr[1|1])y > 0;$$
$$-(3Pr[0|0] + 2Pr[2|0])x + (3Pr[3|0] + 2Pr[1|0])y > 0;$$

All 4 inequalities are satisfied if and only if

$$3Pr[3|1] + 2Pr[1|1] < 3Pr[0|1] + 2Pr[2|1];$$
$$3Pr[0|0] + 2Pr[2|0] < 3Pr[3|0] + 2Pr[1|0];$$

which is impossible.

*No NE where one agent always reports 1, n agents report according to $s^{lie}$ and $3 - n$ agents report honestly.* Clearly, when all 3 agents report honestly, the agent always reporting 1 has the incentive to deviate and report 0 after observing low quality. Consider the strategy profile $s = (s^{pos}, 2 \times \bar{s}, s^{lie})$ where one agent reports according to $s^{pos}$, two agents report honestly and one agent reports according to $s^{lie}$. For the liar, $s^{lie}$ is an equilibrium iff:

$$-\hat{Pr}[0|0]x + \hat{Pr}[1|0]y > 0;$$
$$\hat{Pr}[0|1]x - \hat{Pr}[1|1]y > 0;$$

where $\hat{Pr}[n|o_i]$ is the probability that $n$ out of the 2 honest reporters will observe 1, given the observation $o_i \in Q_2$. By Lemma 3.1, we have $\frac{\hat{Pr}[1|1]}{\hat{Pr}[0|1]} > \frac{\hat{Pr}[1|0]}{\hat{Pr}[0|0]}$, so the above inequations cannot hold simultaneously.

Consider the strategy profile $s = (s^{pos}, \bar{s}, 2 \times s^{lie})$ where one agent reports according to $s^{pos}$, one agent reports honestly and two agents report according to $s^{lie}$. The agent reporting honestly, does so iff:

$$\hat{Pr}[2|0]x - \hat{Pr}[0|0]y > 0;$$
$$-\hat{Pr}[2|1]x + \hat{Pr}[0|1]y > 0;$$

where $\hat{Pr}[n|o_i]$ is the probability that $n$ out of the 2 liars observe 1, given the observation $o_i \in Q_2$. This is impossible since by Lemma 3.1 $\frac{\hat{Pr}[0|0]}{\hat{Pr}[2|0]} > \frac{\hat{Pr}[0|1]}{\hat{Pr}[2|1]}$.

Similar techniques can be used to prove that no strategy profile $\left(s^{neg}, n \times s^{lie}, (3-n) \times \bar{s}\right)$ or $\left(s^{neg}, s^{pos}, n \times s^{lie}, (2-n) \times \bar{s}\right)$ can be NE. Therefore, the only constraint (besides the incentive-compatibility constraints) acting on the payments $x$ and $y$ is intended to prevent the all lying equilibrium. $x$ and $y$ take exactly the values described by Proposition 4.2.

## Appendix C. Proof of Proposition 4.6

Consider the strategy profile $s = \left(\bar{n} \times \bar{s}, (N - \bar{n}) \times s^{lie}\right) \in \tilde{S}$ where $\bar{n} \geq 1$ agents report honestly, and the others always lie. If $s$ were a NE, an honest reporter must expect a higher payment by reporting the truth, while a liar must expect a higher payment by lying. Consider an honest reporter observing 0. She will report a negative signal if and only if $Pr[r_{-i} = 1|0]x > Pr[r_{-i} = N - 2|0]y$, where $Pr[r_{-i} = 1|0]$ and $Pr[r_{-i} = N - 2|0]$ are the probabilities that exactly 1, respectively $N - 2$ of the remaining $N - 1$ agents report positive signals. Exactly one of the other agents reports a positive signal when:

- all but one of the other honest reporters observes low quality, and all liars observe high quality, or

- all honest reporters observe low quality, and all but one of the liars observe high quality.

$$Pr[r_{-i} = 1|0] = \sum_{\theta \in \Theta} Pr[\theta|0] \binom{\bar{n}-1}{1} Pr[1|\theta] Pr[0|\theta]^{\bar{n}-2} \binom{N-\bar{n}}{N-\bar{n}} Pr[1|\theta]^{N-\bar{n}} +$$

$$\sum_{\theta \in \Theta} Pr[\theta|0] \binom{\bar{n}-1}{0} Pr[0|\theta]^{\bar{n}-1} \binom{N-\bar{n}}{N-\bar{n}-1} Pr[1|\theta]^{N-\bar{n}-1} Pr[0|\theta]$$

$$= \frac{(\bar{n}-1)!(N-\bar{n}+1)!}{(N-1)!} Pr[N-\bar{n}+1|0] + \frac{(\bar{n})!(N-\bar{n})!}{(N-1)!} Pr[N-\bar{n}-1|0]$$

$$= \frac{(\bar{n}-1)!(N-\bar{n})!}{(N-1)!} \Big( (N-\bar{n}+1) Pr[N-\bar{n}+1|0] + \bar{n} Pr[N-\bar{n}-1|0] \Big);$$

Similarly,

$$Pr[r_{-i} = N - 2|0] = \frac{(\bar{n}-1)!(N-\bar{n})!}{(N-1)!} \Big( (N-\bar{n}+1) Pr[\bar{n}-2|0] + \bar{n} Pr[\bar{n}|0] \Big);$$

Hence the honest reporter has the incentive to truthfully submit a negative report if and only if:

$$(N-\bar{n}+1) \Big( Pr[N-\bar{n}+1|0]x - Pr[\bar{n}-2|0]y \Big) + \bar{n} \Big( Pr[N-\bar{n}-1|0]x - Pr[\bar{n}|0]y \Big) > 0;$$

On the other hand, the honest reporter will submit a positive report after observing a high quality signal if and only if:

$$(N-\bar{n}+1) \Big( Pr[\bar{n}-2|1]y - Pr[N-\bar{n}+1|1]x \Big) + \bar{n} \Big( Pr[\bar{n}|1]y - Pr[N-\bar{n}-1|1]x \Big) > 0;$$

Exactly the same reasoning leads to the following two inequations for the liar:

$$(N-\bar{n}) \Big( Pr[\bar{n}-1|0]y - Pr[N-\bar{n}|0]x \Big) + (\bar{n}+1) \Big( Pr[\bar{n}+1|0]y - Pr[N-\bar{n}-2|0]x \Big) > 0;$$

$$(N-\bar{n}) \Big( Pr[N-\bar{n}|1]x - Pr[\bar{n}-1|1]y \Big) + (\bar{n}+1) \Big( Pr[N-\bar{n}-2|1]x - Pr[\bar{n}+1|1]y \Big) > 0;$$

There exist $x$ and $y$ such that the four inequalities are satisfied in the same time only if:

$$\frac{(N-\bar{n}+1)Pr[\bar{n}-2|0] + \bar{n}Pr[\bar{n}|0]}{(N-\bar{n}+1)Pr[N-\bar{n}+1|0] + \bar{n}Pr[N-\bar{n}-1|0]} < \frac{(N-\bar{n}+1)Pr[\bar{n}-2|1] + \bar{n}Pr[\bar{n}|1]}{(N-\bar{n}+1)Pr[N-\bar{n}+1|1] + \bar{n}Pr[N-\bar{n}-1|1]}$$

$$\frac{(N-\bar{n})Pr[\bar{n}-1|1] + (\bar{n}+1)Pr[\bar{n}+1|1]}{(N-\bar{n})Pr[N-\bar{n}|1] + (\bar{n}+1)Pr[N-\bar{n}-2|1]} < \frac{(N-\bar{n})Pr[\bar{n}-1|0] + (\bar{n}+1)Pr[\bar{n}+1|0]}{(N-\bar{n})Pr[N-\bar{n}|0] + (\bar{n}+1)Pr[N-\bar{n}-2|0]}$$

or equivalently:

$$\frac{(N-\bar{n}+1)Pr[\bar{n}-2|0] + \bar{n}Pr[\bar{n}|0]}{(N-\bar{n}+1)Pr[\bar{n}-2|1] + \bar{n}Pr[\bar{n}|1]} < \frac{(N-\bar{n}+1)Pr[N-\bar{n}+1|0] + \bar{n}Pr[N-\bar{n}-1|0]}{(N-\bar{n}+1)Pr[N-\bar{n}+1|1] + \bar{n}Pr[N-\bar{n}-1|1]}$$

$$\frac{(N-\bar{n})Pr[\bar{n}-1|1] + (\bar{n}+1)Pr[\bar{n}+1|1]}{(N-\bar{n})Pr[\bar{n}-1|0] + (\bar{n}+1)Pr[\bar{n}+1|0]} < \frac{(N-\bar{n})Pr[N-\bar{n}|1] + (\bar{n}+1)Pr[N-\bar{n}-2|1]}{(N-\bar{n})Pr[N-\bar{n}|0] + (\bar{n}+1)Pr[N-\bar{n}-2|0]}$$

However, one can show that:

$$\frac{(N-\bar{n})Pr[N-\bar{n}|1] + (\bar{n}+1)Pr[N-\bar{n}-2|1]}{(N-\bar{n})Pr[N-\bar{n}|0] + (\bar{n}+1)Pr[N-\bar{n}-2|0]} < \frac{(N-\bar{n}+1)Pr[\bar{n}-2|0] + \bar{n}Pr[\bar{n}|0]}{(N-\bar{n}+1)Pr[\bar{n}-2|1] + \bar{n}Pr[\bar{n}|1]}$$

and

$$\frac{(N-\bar{n}+1)Pr[N-\bar{n}+1|0] + \bar{n}Pr[N-\bar{n}-1|0]}{(N-\bar{n}+1)Pr[N-\bar{n}+1|1] + \bar{n}Pr[N-\bar{n}-1|1]} < \frac{(N-\bar{n})Pr[\bar{n}-1|1] + (\bar{n}+1)Pr[\bar{n}+1|1]}{(N-\bar{n})Pr[\bar{n}-1|0] + (\bar{n}+1)Pr[\bar{n}+1|0]}$$

which means that the honest reporter and the liar cannot both believe that their strategies are optimal (given the strategies of the other agents).

Consider the strategy profile $s = (s^{neg}, \bar{n} \times \bar{s}, N - \bar{n} - 1 \times s^{lie}) \in \tilde{S}$ where one agent always reports 0, $\bar{n} \geq 1$ agents report honestly, and $N - \bar{n} - 1 \geq 1$ agents always lie. An honest reporter and a liar both believe that $s$ is a NE if and only if:

$$
\begin{aligned}
\left( \bar{n} \bar{P} r[N - \bar{n} - 2|0] + (N - \bar{n}) \bar{P} r[N - \bar{n}|0] \right) x - \bar{P} r[\bar{n} - 1|0] y &> 0 \\
-\left( \bar{n} \bar{P} r[N - \bar{n} - 2|1] + (N - \bar{n}) \bar{P} r[N - \bar{n}|1] \right) x + \bar{P} r[\bar{n} - 1|1] y &> 0 \\
-\left( (\bar{n} + 1) \bar{P} r[N - \bar{n} - 3|0] + (N - \bar{n} - 1) \bar{P} r[N - \bar{n} - 1|0] \right) x + \bar{P} r[\bar{n}|0] y &> 0 \\
\left( (\bar{n} + 1) \bar{P} r[N - \bar{n} - 3|1] + (N - \bar{n} - 1) \bar{P} r[N - \bar{n} - 1|1] \right) x - \bar{P} r[\bar{n}|1] y &> 0
\end{aligned}
\tag{17}
$$

where $\bar{P} r[j|o_i]$ is the probability that $j$ out of $N - 2$ agents observe high quality signals, given the observation $o_i$.

Nevertheless,

$$
\frac{\bar{P} r[\bar{n}|0]}{\bar{P} r[\bar{n}|1]} < \frac{\bar{P} r[\bar{n} - 1|0]}{\bar{P} r[\bar{n} - 1|1]}
$$

and

$$
\frac{\bar{n} \bar{P} r[N - \bar{n} - 2|0] + (N - \bar{n}) \bar{P} r[N - \bar{n}|0]}{\bar{n} \bar{P} r[N - \bar{n} - 2|1] + (N - \bar{n}) \bar{P} r[N - \bar{n}|1]} < \frac{(\bar{n} + 1) \bar{P} r[N - \bar{n} - 3|0] + (N - \bar{n} - 1) \bar{P} r[N - \bar{n} - 1|0]}{(\bar{n} + 1) \bar{P} r[N - \bar{n} - 3|1] + (N - \bar{n} - 1) \bar{P} r[N - \bar{n} - 1|1]}
$$

which means that the inequalities in (17) can never be simultaneously satisfied.

Using exactly the same technique one can show that:

- $s = (s^{pos}, \bar{n} \times \bar{s}, (N - \bar{n} - 1) \times s^{lie}) \in \tilde{S}$ where one agent always reports 0, $\bar{n} \geq 1$ agents report honestly, and $N - \bar{n} - 1 \geq 1$ agents always lie is not a NE;

- $s = (s^{neg}, s^{pos}, \bar{n} \times \bar{s}, (N - \bar{n} - 2) \times s^{lie}) \in \tilde{S}$ where one agent always reports 0, one agent always reports 1, $\bar{n} \geq 0$ agents report honestly, and $N - \bar{n} - 1 \geq 0$ agents always lie is not a NE.

## Appendix D. Extending the results to n-ary feedback

We have assumed so far that agent can observe and report only two signals: 0 or 1. The framework can be extended to n-ary feedback by imposing supplementary constraints on the design problems. For example, let's assume the set of quality signals (and feedback reports) contains $M$ elements: $Q = \{q_1, q_2, \ldots q_M\}$. The incentive compatibility constraints equivalent to (4) become:

$$
\sum_{r_{-i} \in Q^{(N-1)}} Pr[r_{-i}|q_j] \left( \tau(q_j, r_{-i}) - \tau(q_k, r_{-i}) \right) \geq \Lambda; \ \forall q_j, q_k \in Q;
$$

where $q_j \in Q$ is the signal actually observed by agent $i$, $q_k \in Q$ is every other lie agent $i$ could report, and $r_{-i} \in Q^{(N-1)}$ is the configuration of the other $N - 1$ reference reports. When compared to (4) three observations become apparent. First, the 2 constraints for the binary feedback case must be replaced by $M(M - 1)$ constraints for the n-ary feedback setting. Second, the 'size' of each constraint grows from a sum of $N$ terms to a sum of $C_{N+M-2}^{M-1}$ terms (all possible unordered sequences of $N-1$ signals drawn from $Q$). Finally, the reward mechanism is defined by $M \cdot C_{N+M-2}^{M-1}$ payments instead of $2N$.

All other constraints will suffer a similar blowup, which makes the design problem for the general n-ary feedback case significantly harder to solve.